\documentclass[12pt]{article}

\font\grande=cmr9.5 scaled \magstep4
\font\medio=cmr9.5 scaled \magstep2
\outer\def\beginsection#1\par{\medbreak\bigskip
      \message{#1}\leftline{\bf#1}\nobreak\medskip
\vskip-\parskip
      \noindent}
\usepackage{graphicx} 
\begin{document}
\bibliographystyle {unsrt}

\titlepage

\begin{flushright}
CERN-PH-TH/2012-218
\end{flushright}

\vspace{10mm}
\begin{center}
{\grande Secondary graviton spectra, second-order correlations}\\
\vspace{1cm}
{\grande and Bose-Einstein enhancement}\\
\vspace{1.5cm}
 Massimo Giovannini
 \footnote{Electronic address: massimo.giovannini@cern.ch}\\
\vspace{1cm}
{{\sl Department of Physics, 
Theory Division, CERN, 1211 Geneva 23, Switzerland }}\\
\vspace{0.5cm}
{{\sl INFN, Section of Milan-Bicocca, 20126 Milan, Italy}}
\vspace*{0.5cm}
\end{center}

\vskip 0.5cm
\centerline{\medio  Abstract}
\vskip 0.2cm
Primary graviton spectra, produced via stimulated emission from an initial  Bose-Einstein distribution, are enhanced for typical scales 
larger than the redshifted thermal wavelength. A mixed state of phonons induces a
secondary graviton spectrum which is hereunder computed in terms of three parameters  (i.e. the number of phonon species, the tensor-to-scalar ratio and the thermal wavelengths of the mixture). The primary and secondary graviton spectra are shown to be sensitive, respectively, to the first-order and second-order correlation properties of the initial quantum mixture so that the semiclassical theory is argued to be generally inadequate in this context. For particular values of the parameters the secondary contribution may turn out to be comparable with the primary spectrum over large-scales. 
\vskip 0.5cm
\noindent
\vspace{5mm}

\vfill
\newpage
\renewcommand{\theequation}{1.\arabic{equation}}
\setcounter{equation}{0}
\section{Introduction}
\label{sec1}
The analysis of the Cosmic Microwave Background (CMB) 
anisotropies and polarization \cite{wmap1,wmap2,ACBAR,QUAD}, of the extended galaxy surveys \cite{LSS1,LSS2} and of  the supernova data sets \cite{SNN1,SNN2} aims at a highly accurate 
statistical determination of the cosmological parameters belonging to a class of simplified scenarios developed around the so-called concordance paradigm also dubbed, sometimes, $\Lambda$CDM 
model\footnote{In the acronym the $\Lambda$ term parametrizes the dark energy component and CDM stands for the cold dark matter component}. In the concordance paradigm the tensor modes are assumed to be absent but there are good reasons to believe in their existence. It is plausible to think that, in the near future, we shall be forced to deal with some sort of  T$\Lambda$CDM scenario \cite{mg1} (with T standing for the tensor modes) where experiments operating at high-frequencies\footnote{Current CMB and large-scale observations are commonly phrased in terms of wavenumbers expressed in units of $\mathrm{Mpc}^{-1}$ while the operational windows of wide-band interferometers (cross-correlated among themselves and with some bar detector) are given in Hz. It is useful to recall, in this respect, that the pivot scale at which the CMB power spectra are assigned corresponds to $k_{\mathrm{p}} = 0.002\, \mathrm{Mpc}^{-1}$ (related to an effective multipole $\ell_{\mathrm{p}} = 30$). In the natural units used throughout this paper (i.e. $\hbar= c= \kappa_{\mathrm{B}} =1$) the frequency $\nu_{\mathrm{p}} = k_{\mathrm{p}}/(2 \pi) \simeq 3.092\times 10^{-18} \,$ Hz. The devices quoted in \cite{WB1,WB2} roughly probe regions between $60$ and $110$ Hz while the ones of Ref. \cite{HF1,HF2} are expected to work in the MHz region.}(such as the wide-band 
interferometers \cite{WB1,WB2} or devices operating at even larger frequencies \cite{HF1,HF2}) and CMB  polarization experiments could be jointly analyzed to probe possible excesses both over large length-scales (i.e. small frequencies) and over comparatively higher frequencies.

The precise nature of the pre-inflationary initial conditions calls for supplementary assumptions \cite{wein} and it is anyway not explained in the framework of the simplistic $\Lambda$CDM paradigm. It is useful to speculate that the spectrum of relic gravitons has a steeper slope at large scales.  This is what happens if thermal corrections to the graviton spectrum are consistently included in the scenario.
The thermal corrections arise if a Bose-Einstein distribution of spin-two fluctuations 
of the geometry  was present around $65$ efolds prior to the end of a conventional inflationary phase (see, e.g. \cite{one,two,twoa}).  The correction to the power spectrum takes a standard form which can be written as \cite{mg2,three,four,five,six,seven,eight}
\begin{equation}
P^{(1)}_{\mathrm{T}}(k,k_{\mathrm{T}}) = \coth{\biggl(\frac{k}{2 k_{\mathrm{T}}}\biggr)} \overline{P}_{\mathrm{T}}(k),
\label{eq1}
\end{equation}
where $k_{\mathrm{T}} = T$ is the comoving thermal wavenumber in units $\hbar = c = \kappa_{\mathrm{B}} = 1$; $\overline{P}_{\mathrm{T}}(k)$ is the spectrum in the absence of thermal corrections so that
the two spectra are not distinguishable when $k \gg k_{\mathrm{T}}$ while the spectral slopes differ for 
wavelenghts $\lambda > \lambda_{\mathrm{T}} = 2\pi/k_{\mathrm{T}}$. In the limit $\lambda \gg \lambda_{\mathrm{T}}$, $P^{(1)}_{\mathrm{T}}(k,k_{\mathrm{T}}) \simeq (k/k_{\mathrm{T}})^{-1} \,\overline{P}_{\mathrm{T}}(k)$.  More general situations can be contemplated if the initial state 
of the gravitons is mixed but not exactly thermal. In all these situations particles are produced by stimulated 
emission as opposed to the spontaneous emission operating when the initial state is the vacuum. For wavenumbers $k \gg k_{\mathrm{T}}$, as a consequence of Eq. (\ref{eq1}), the stimulated and spontaneous emissions lead, in practice, to the same results. 

The idea pursued in this paper is, in short, the following. If the initial state of the gravitons is mixed, then
it is also plausible that all the other fields (both conformally and non conformally coupled)  will 
be in  a mixed state characterized either by the same or by a similar comoving temperature. 
Among all the possible fields with different spins, particular importance should be attributed to the non-conformally coupled species.  
To address this broad problem it is useful to consider the simplest 
non-trivial situation where the initial state of the fluctuations 
is given as the direct product of the quantum state of the gravitons and of the other quantum states 
of the matter fields which will be assumed to consist of a collection of phonons (i.e. non-conformally 
coupled scalar fields):
\begin{equation}
| \Psi \rangle = |\psi_{\mathrm{gravitons}}\rangle\, \otimes\, |\phi_{\mathrm{phonons}}\rangle \otimes\,....
\label{eq2}
\end{equation}
where the ellipses stand for other matter fields eventually present in the system. Both the gravitons 
and the phonons will be characterized by a density matrix whose statistical weights define 
the correlation properties of the initial state, as it will be specifically discussed hereunder. 

The secondary graviton spectra induced by a mixture 
of phonons are the main theme of the present analysis which may serve as a useful 
approximation to more realistic (but also necessarily more complicated) situations. The secondary graviton spectra, unlike their primary counterpart, are sensitive to the second-order correlation properties of the initial quantum mixture so that the semiclassical theory is argued to be generally inadequate. In section \ref{sec2}
the quantum treatment of the anisotropic stress will be outlined. In section \ref{sec3} the discussion focuses on the relation between the secondary spectrum and the correlation properties of the initial state. In section
\ref{sec4} the secondary spectrum is deduced in the realistic situation. Section \ref{sec5} contains the concluding considerations and some perspectives for future developments 
in the light of forthcoming satellite experiments. 

\renewcommand{\theequation}{2.\arabic{equation}}
\setcounter{equation}{0}
\section{Quantum anisotropic stresses}
\label{sec2}
A simple way to pin down the primary contribution to the graviton spectrum 
is to consider the limit where  the gravitons are decoupled from the phonons but not from the initial mixed state. The secondary contribution depends instead on the contribution of the phonons to the total anisotropic stress which must be 
viewed, for the present purposes, as a field operator.
Defining three mutually orthogonal directions as $\hat{k}_{i} = k_{i}/|\vec{k}|$,  $\hat{m}_{i} = m_{i}/|\vec{m}|$ and $\hat{n} =n_{i}/|\vec{n}|$, the two polarizations of the gravitons in a conformally flat background are:
 \begin{equation}
 e_{ij}^{(\oplus)}(\hat{k}) = (\hat{m}_{i} \hat{m}_{j} - \hat{n}_{i} \hat{n}_{j}), \qquad 
 e_{ij}^{(\otimes)}(\hat{k}) = (\hat{m}_{i} \hat{n}_{j} + \hat{n}_{i} \hat{m}_{j}).
 \label{ST0}
 \end{equation}
From Eq. (\ref{ST0}) it follows that $e_{ij}^{(\lambda)}\,e_{ij}^{(\lambda')} = 2 \delta_{\lambda\lambda'}$. The sum over the polarization gives instead:
\begin{equation}
\sum_{\lambda} e_{ij}^{(\lambda)}(\hat{k}) \, e_{m n}^{(\lambda)}(\hat{k}) = 
 \biggl[p_{m i}(\hat{k}) p_{n j}(\hat{k}) + p_{m j}(\hat{k}) p_{n i}(\hat{k}) - p_{i j}(\hat{k}) p_{m n}(\hat{k}) \biggr],
\label{ST0B} 
\end{equation}
where $p_{ij}(\hat{k}) = (\delta_{i j} - \hat{k}_{i} \hat{k}_{j})$  is the traceless projector. The primary spectra of the relic gravitons can be deduced from the following action:
\begin{equation}
S = \frac{1}{8 \ell_{\mathrm{P}}^2} \int d^{4} x \sqrt{- \overline{g}}\overline{g}^{\alpha\beta}
\partial_{\alpha} \, h_{i}^{j} \, \partial_{\beta} h_{j}^{i},
\label{action1}
\end{equation}
where $\overline{g}_{\mu\nu}$ denotes a conformally flat background metric\footnote{Greek letters are used to denote four-dimensional indices; lowercase Latin characters denote spatial indices; uppercase Latin characters 
denote the different phonon indices.} and $\ell_{\mathrm{P}} = \sqrt{8 \pi G} = 1/\overline{M}_{\mathrm{P}}$. In what follows we shall distinguish (especially in the last section) between $\overline{M}_{\mathrm{P}}$ and $M_{\mathrm{P}} = 1.22\times 10^{19} \mathrm{GeV} = \sqrt{8 \pi} \,\,\overline{M}_{\mathrm{P}}$. In the action (\ref{action1}) $h_{ij}$ denote the traceless and divergenceless modes of the geometry written in the form $g_{\mu\nu}(\vec{x},\tau) = 
\overline{g}_{\mu\nu}(\tau) + \delta_{\mathrm{t}}g_{\mu\nu}(\vec{x},\tau)$ where 
$\delta_{\mathrm{t}} g_{i j } = - a^2 \, h_{ij}$ and $\delta g^{i j} = h^{ij}/a^2$.
By taking the functional variation of the action (\ref{action1}) with respect to $h_{i}^{j}$ the equation of motion reads
\begin{equation}
h_{ij}^{\prime\prime} + 2 {\mathcal H} h_{ij}^{\prime} - \nabla^2 h_{ij} = 0,
\label{mot1A}
\end{equation}
where $\overline{g}_{\mu\nu} = a^2(\tau)\eta_{\mu\nu}$.  In Eq. (\ref{mot1A}) 
the prime denotes a derivation with respect to the conformal time 
coordinate $\tau$ and ${\mathcal H} = (\ln{a})' = a\,H$ where $H$ is the Hubble rate. 
\subsection{Primary spectra}
In terms of the two polarizations of Eq. (\ref{ST0}) the mode expansion of the field operator $\hat{h}_{ij}(\vec{x},\tau)$ becomes\footnote{The superscript in Eq. (\ref{ST1}) stands for the primary contribution which 
is the one  obtainable by setting the phonon contribution to zero.}
 \begin{equation}
 \hat{h}^{(1)}_{ij}(\vec{x},\tau)= \frac{\sqrt{2} \ell_{\mathrm{P}}}{(2\pi)^{3/2} a(\tau)} \sum_{\lambda}
  \int d^{3}k\,\, e_{ij}^{(\lambda)}(\hat{k})\, 
 \biggl[ \hat{a}_{\vec{k},\lambda} \,f_{k,\lambda}(\tau) e^{- i \vec{k} \cdot \vec{x}} +  \hat{a}_{\vec{k},\lambda}^{\dagger}\, f_{k,\lambda}^{*}(\tau) 
 e^{ i \vec{k} \cdot \vec{x}}\biggr],
 \label{ST1}
 \end{equation}
 where $\hat{a}_{\vec{k},\lambda}$ and $f_{k,\lambda}(\tau)$ obey, respectively, 
 \begin{equation}
 [\hat{a}_{\vec{k},\lambda}, \hat{a}^{\dagger}_{\vec{p},\lambda'}] = \delta^{(3)}(\vec{k} - \vec{p}) \delta_{\lambda\lambda'}, \qquad f_{k\,\lambda}'' + \biggl[ k^2 - \frac{a''}{a} \biggr] f_{k\, \lambda} =0. 
 \label{ST1a}
 \end{equation}
The sum over $\lambda$ appearing in Eq. (\ref{ST1}) runs obviously over the 
two polarizations defined in Eq. (\ref{ST0}).
The field operator of Eq. (\ref{ST1}) can also be represented, in Fourier space, as
\begin{eqnarray}
\hat{h}^{(1)}_{ij}(\vec{q},\tau) &=& \frac{1}{(2\pi)^{3/2}} \int d^{3} x  \, \hat{h}^{(1)}_{ij}(\vec{x},\tau) \,e^{i \vec{q}\cdot\vec{x}}
\nonumber\\
&=& \frac{\sqrt{2} \ell_{\mathrm{P}}}{a(\tau)} 
\sum_{\lambda} \biggl[ \hat{a}_{\vec{q},\lambda} \,f_{k,\lambda}(\tau) e_{ij}^{(\lambda)}(\hat{q})+  \hat{a}_{- \vec{q},\lambda}^{\dagger}\, f_{k,\lambda}^{*}(\tau) e_{ij}^{(\lambda)}(-\hat{q})\biggr].
\label{ST2}
\end{eqnarray}
To compute the primary spectrum we need to calculate the 
expectation value
\begin{eqnarray}
&& \langle \Psi | \hat{h}^{(1)}_{ij}(\vec{q}_{1}, \tau) \, \hat{h}_{m n}^{(1)\, \dagger}(\vec{q}_{2}, \tau') |  \Psi\rangle
\nonumber\\
&& = \langle \psi | \hat{h}^{(1)}_{ij}(\vec{q}_{1}, \tau) \, \hat{h}_{m n}^{(1)\, \dagger}(\vec{q}_{2}, \tau') |  \psi\rangle 
= \mathrm{Tr}\bigg[ \hat{\rho}_{\mathrm{gravitons}}\, \hat{h}^{(1)}_{ij}(\vec{q}_{1}, \tau) \, \hat{h}_{m n}^{(1)\, \dagger}(\vec{q}_{2}, \tau') \biggr].
\label{ST2a}
\end{eqnarray}
The first equality of Eq. (\ref{ST2a}) follows from Eq. (\ref{eq2}) while the second equality prescribes that the expectation value must be computed by using the appropriate density matrix for the gravitons. We 
postpone to section \ref{sec3} a more specific discussion of the density matrix itself. For the results of the present 
section the form of the density matrix is immaterial: what really matters is only the averaged multiplicity of the initial state and not its dispersion.  The expectation value defined in Eq. (\ref{ST2a}) can be written, in more explicit terms, using Eq. (\ref{ST2}):
\begin{eqnarray}
&& \frac{2 \ell_{\mathrm{P}}^2}{a^2(\tau)} \sum_{\lambda} \sum_{\lambda'} 
\biggl\{ f_{q_{1}\, \lambda}(\tau) f^{*}_{q_{2}\, \lambda'}(\tau')  e_{ij}^{(\lambda)}(\hat{q}_{1})e_{mn}^{(\lambda')}(\hat{q}_{2})  \langle
\hat{a}_{\vec{q}_{1},\lambda} \hat{a}^{\dagger}_{\vec{q}_{2},\lambda'}\rangle 
\nonumber\\
&& + f_{q_{1}\, \lambda}^{*}(\tau) f_{q_{2}\, \lambda'}(\tau')  e_{ij}^{(\lambda)}(- \hat{q}_{1})e_{mn}^{(\lambda')}(- \hat{q}_{2})  \langle
\hat{a}^{\dagger}_{-\vec{q}_{1},\lambda} \hat{a}_{-\vec{q}_{2},\lambda'} \rangle \biggr\}.
\label{ST3}
\end{eqnarray}
Using the Kroeneker deltas over the polarizations and the Dirac 
deltas over the momenta Eq. (\ref{ST3}) becomes, for $\tau\to \tau'$ 
\begin{equation}
\langle \psi | \hat{h}^{(1)}_{ij}(\vec{q}_{1}, \tau) \, \hat{h}_{m n}^{(1)\, \dagger}(\vec{q}_{2}, \tau) |  \psi\rangle = \frac{8 \ell_{\mathrm{P}}^2}{a^2(
\tau)} \, {\mathcal S}_{i j m n}(\hat{q}_{1})\, |f_{q_{1}}(\tau)|^2\, [2 \overline{n}(\vec{q}_{1}) +1] \delta^{(3)}(\vec{q_{1}}- \vec{q}_{2}).
\label{ST4}
\end{equation}
In Eq. (\ref{ST4}) the following auxiliary sum  ${\mathcal S}_{i j m n}(\hat{k}) =  \sum_{\lambda} e_{ij}^{(\lambda)}(\hat{k}) \, e_{m n}^{(\lambda)}(\hat{k})/4$ has been defined; recalling Eq. (\ref{ST2}), $\overline{n}(q_{1})$ denotes the averaged multiplicity of the initial state
\begin{equation}
\overline{n}(q_{1})
\delta_{\lambda\lambda'} \, \delta^{(3)}(\vec{q}_{1} - \vec{q}_{2})
= \langle \psi| \hat{a}^{\dagger}_{\vec{q}_{1},\lambda} \hat{a}_{\vec{q}_{2},\lambda'} |\psi \rangle  \equiv 
\mathrm{Tr}\bigg[ \hat{\rho}_{\mathrm{gravitons}}\,  \hat{a}^{\dagger}_{\vec{q}_{1},\lambda} \hat{a}_{\vec{q}_{2},\lambda'} \biggr].
\label{ST5}
\end{equation}
In Eq. (\ref{ST4}) the mode functions of each polarization obey the same equation, i.e. where $f_{q_{1}\,\otimes}= f_{q_{1},\,\oplus} \equiv f_{q_{1}}$. Recalling now that, from Eq. (\ref{ST0B}),  
${\mathcal S}_{i j i j}(\hat{k}) = 1$, the primary power spectrum $P^{(1)}_{\mathrm{T}}(q,\tau)$ arises 
from the first-order correlation between the graviton field operators
\begin{eqnarray}
\langle \psi | \hat{h}^{(1)}_{ij}(\vec{q}, \tau) \, \hat{h}_{ij}^{(1)\, \dagger}(\vec{p}, \tau) |  \psi\rangle &=& \frac{2\pi^2}{q^3} P^{(1)}_{\mathrm{T}}(q,\tau)
\, \delta^{(3)}(\vec{q} - \vec{p}), 
\label{ST6A}\\
P_{\mathrm{T}}^{(1)}(q,\tau) &=& \frac{4\ell_{\mathrm{P}}^2\,q^3}{\pi^2 a^2(\tau)} 
|f_{q}(\tau)|^2 \, [2 \overline{n}(q) + 1].
\label{ST6B}
\end{eqnarray}
In the limit $\overline{n}(q) \to 0$, using the results the mode functions calculated during a quasi-de Sitter stage (see, e.g. section \ref{sec4}  Eqs. (\ref{BE2})--(\ref{BE3})), Eq. (\ref{ST6B}) gives exactly the properly normalized power spectrum of the tensor modes of the geometry of inflationary origin. 
\subsection{Energy densities}
The energy-momentum tensor of the gravitons can be derived from the action (\ref{action1}) by varying with respect to $\overline{g}^{\alpha\beta}$. As argued by Ford and Parker \cite{ford} the resulting expression is a reasonable (but not unique) quantity to use for the calculation of the spectral energy density of the gravitons; the energy-momentum tensor is given, in this case, by:
\begin{equation}
{\mathcal T}_{\mu}^{\nu} = \frac{1}{4\ell_{\mathrm{P}}^2}\biggl[ \partial_{\mu} h_{ij} \partial^{\nu} h^{ij} - \frac{1}{2} \delta_{\mu}^{\nu} 
g^{\alpha\beta} \partial_{\alpha} h_{ij} \partial_{\beta} h^{ij} \biggr].
\label{enmom1}
\end{equation}
The energy density of the gravitons is therefore obtained by averaging the  $(00)$ component of Eq. (\ref{enmom1}) over the appropriate (mixed) state defined introduced in Eq. (\ref{eq2}) and 
already mentioned earlier in this section, i.e.  $\rho_{\mathrm{GW}} = \langle \psi | {\mathcal T}_{0}^{0} |\psi \rangle$. Defining $g_{q} = f_{q}'$ and recalling Eq. (\ref{ST5}) the energy density of the gravitons is
\begin{equation}
\rho_{\mathrm{GW}}
= \frac{1}{a^4} \int d\ln{q} \frac{q^3}{2\pi^2} \biggl\{ |g_{q}(\tau)|^2 + ( q^2 + {\mathcal H}^2) |f_{q}(\tau)|^2
- {\mathcal H}[ f_{q}^{*}(\tau)g_{q}(\tau) + f_{q}(\tau) g_{q}^{*}(\tau)] \biggr\} [ 2 \overline{n}(q) +1].
\label{enmom1b}
\end{equation}
By measuring Eq. (\ref{enmom1b}) in units of the critical energy density 
$\rho_{\mathrm{crit}}$  the critical fraction of energy density stored in the gravitons is obtained:
\begin{equation}
\Omega_{\mathrm{GW}}(q,\tau) = \frac{1}{\rho_{\mathrm{crit}}} \frac{d\, \rho_{\mathrm{GW}}}{d \ln{q}},\qquad \rho_{\mathrm{GW}} = \langle \psi | {\mathcal T}_{0}^{0} |\psi \rangle, \qquad \rho_{\mathrm{crit}} = \frac{3 {\mathcal H}^2}{a^2\, \ell_{\mathrm{P}}^2}.
\label{enmom1c}
\end{equation}
Using Eq. (\ref{enmom1b}),  Eq. (\ref{enmom1c}) becomes
\begin{equation}
\Omega_{\mathrm{GW}}(q,\tau) =\frac{q^2 P_{\mathrm{T}}^{(1)}(q,\tau)}{ 24 {\mathcal H}^2}  \biggl\{ 1 + \frac{|g_{q}(\tau)|^2 
+ {\mathcal H}^2 |f_{q}(\tau)|^2 - {\mathcal H}[ f_{q}^{*}(\tau) g_{q}(\tau) + 
f_{q}(\tau) g_{q}^{*}(\tau)]}{q^2 |f_{q}(\tau)|^2} \biggr\}.
\label{enmom2}
\end{equation}
Instead of deriving the energy density from Eq. (\ref{enmom1}), it is 
equally plausible to use the energy-momentum pseudo-tensor \cite{landau1} appropriately generalized to curved backgrounds \cite{landau2,landau3}. By defining the energy-momentum pseudo-tensor from the second-order variation of the Einstein tensor,  the expression for $\rho_{\mathrm{GW}}$ is given by \cite{mg3,suzhang}
\begin{equation}
\rho_{\mathrm{GW}} =  \frac{1}{a^2 \ell_{\rm P}^2} \biggl[ {\mathcal H} 
\langle \psi | \partial_{\tau} h_{k\ell } h^{k\ell} |\psi \rangle + \frac{1}{8} ( \langle \psi | \partial_{m} h_{k\ell} \partial^{m} h^{k\ell} | \psi \rangle + 
\langle \psi |\partial_{\tau} h_{k\ell} \partial_{\tau} {h^{k\ell}} | \psi \rangle)\biggr],
\label{enmom3}
\end{equation}
implying that $\Omega_{\mathrm{GW}}(q,\tau)$ is now given by
\begin{equation}
\Omega_{\mathrm{GW}}(q,\tau) = \frac{q^2 P_{\mathrm{T}}^{(1)}(q,\tau) }{ 24 {\mathcal H}^2} \biggl\{ 1 + \frac{|g_{q}(\tau)|^2 -
7 {\mathcal H}^2 |f_{q}(\tau)|^2 +3 {\mathcal H}[ f_{q}^{*}(\tau) g_{q}(\tau) + 
f_{q}(\tau) g_{q}^{*}(\tau)]}{q^2 |f_{q}(\tau)|^2} \biggr\}.
\label{enmom4}
\end{equation}
The difference between  Eq. (\ref{enmom2}) and Eq. (\ref{enmom4})
only appears in the correction to the leading results. By expanding Eqs. 
(\ref{enmom2}) and (\ref{enmom4}) in the limit ${\mathcal H}/q\ll 1$ the 
results are, respectively: 
\begin{equation}
\Omega_{\mathrm{GW}}(q,\tau) = \frac{q^2 P_{\mathrm{T}}^{(1)}(q,\tau) }{ 12  {\mathcal H}^2} \biggl[1 + 
\frac{{\mathcal H}^2}{2 q^2} \biggr],\qquad 
\Omega_{\mathrm{GW}}(q,\tau) = \frac{q^2 P_{\mathrm{T}}^{(1)}(q,\tau) }{ 12 {\mathcal H}^2} \biggl[1 -
\frac{7{\mathcal H}^2}{2 q^2} \biggr],
\label{enmom5}
\end{equation}
where we used that\footnote{The following two statements are derived 
from the approximate solution of the equation for the mode function 
reported in Eq. (\ref{ST1a}).} 
$g_{q}(\tau) \simeq \pm i q f_{q}(\tau)$  for ${\mathcal H}/q\ll 1$.
 In the opposite limit (i.e. $q/{\mathcal H} \ll 1$) the approximate 
 solution of the mode function reads 
 \begin{equation}
 f_{q}(\tau) \simeq A(q) a(\tau) + B(q) a(\tau) \int^{\tau} \frac{d\tau'}{a^2(\tau')}, \qquad g_{q}(\tau) \simeq {\mathcal H} f_{q} + \frac{B(q)}{a(\tau)}.
 \end{equation}
For instance, during a quasi-de Sitter stage of expansion the term
going as $B(q)/a(\tau)$ is suppressed and
 the expressions of $\Omega_{\mathrm{GW}}(q,\tau)$ coincide 
 up to terms of order $(q/{\mathcal H})^4$ as it can be checked by inserting 
 $g_{q}(\tau) \simeq {\mathcal H} f_{q}(\tau)$ into the general expressions of Eqs. (\ref{enmom2}) and (\ref{enmom4}).
 
\subsection{Anisotropic stresses}
The spectra obtained so far hold provided the induced anisotropic stress is strictly vanishing. In the opposite 
case Eq. (\ref{mot1A}) inherits a source term; by perturbing the Einstein equations and by focussing 
on the modes which are traceless and divergenceless we shall have 
\begin{equation}
\delta_{\mathrm{t}} R_{i}^{j} = \ell_{\mathrm{P}}^2 \Pi_{i}^{j}, \qquad \delta_{\mathrm{t}} R_{i}^{j} = - \frac{1}{2 a^2} \biggl[ 
h_{i}^{j\,\,\prime\prime} + 2 {\mathcal H} h_{i}^{j\,\,\prime} - \nabla^2 h_{i}^{j} \biggr],
\label{mot2A} 
\end{equation} 
where $\delta_{\mathrm{t}} R_{i}^{j}$ denotes the fluctuation of the Ricci tensor and where 
$\Pi_{i}^{j}$ must be both divergenceless and traceless, i.e.
\begin{equation}
\hat{h}_{i}^{j\,\,\prime\prime} + 2 {\mathcal H} \hat{h}_{i}^{j\,\,\prime} - \nabla^2 \hat{h}_{i}^{j}  = - 2\, a^2 \, \ell_{\mathrm{P}}^2 \hat{\Pi}_{i}^{j}, \qquad 
\hat{\Pi}_{i}^{i} = \partial_{j} \, \hat{\Pi}_{i}^{j},
\label{mot3A}
\end{equation} 
where the classical fields have been promoted to the status of quantum operators (i.e. $h_{ij} \to \hat{h}_{ij}$ and $ \Pi_{i}^{j} \to \hat{\Pi}_{i}^{j}$). The problem we ought to address is to obtain an explicit form of the anisotropic stress $\hat{\Pi}_{ij}$ in Fourier space. By taking the Fourier transform of both sides of Eq. (\ref{mot3A}) we get
\begin{equation}
\hat{h}_{ij}'' + 2 {\mathcal H} \hat{h}_{ij} + q^2 \hat{h}_{ij} = -2 \ell_{\mathrm{P}}^2  a^2(\tau) \hat{\Pi}_{ij}(\vec{q},\tau),
\label{phon7a}
\end{equation}
where $\hat{\Pi}_{ij}(\vec{q},\tau)$ can be appropriately expanded in on the basis of the two polarizations 
of the graviton defined in Eq. (\ref{ST0}):
\begin{equation}  
\hat{\Pi}_{ij}(\vec{q},\tau) = \sum_{\lambda} \, e^{(\lambda)}_{ij}(\hat{q}) \, \hat{\Pi}_{\lambda}(\hat{q},\tau) = e^{\oplus}_{ij}(\hat{q})\, \hat{\Pi}_{\oplus}(\vec{q},\tau) +
e^{\otimes}_{ij}(\hat{q})\hat{\Pi}_{\otimes}(\vec{q},\tau), 
\label{anis1a}
\end{equation}
where 
\begin{equation}
\hat{\Pi}_{\oplus}(\vec{q},\tau) = \frac{1}{2} \sum_{A=1}^{{\mathcal N}} e^{m n}_{\oplus}(\hat{q})\, \hat{\pi}_{m n}^{(A)}(\vec{q},\tau), \qquad \hat{\Pi}_{\otimes}(\vec{q},\tau) = \frac{1}{2} \sum_{A=1}^{{\mathcal N}} e^{m n}_{\otimes}(\hat{q}) \,\hat{\pi}_{m n}^{(A)}(\vec{q},\tau).
\label{anis1b}
\end{equation}
The rank-to tensor $\pi_{m n}^{(A)}(\vec{q},\tau)$ appearing in Eq. (\ref{anis1b}) is in fact a convolution of two field operators:
\begin{eqnarray}
&& \hat{\pi}^{(A)}_{m n}(\vec{q},\tau) = \frac{1}{(2\pi)^{3/2}\, a^2(\tau)} \, \int d^{3} p\, C^{(A)}_{m n}(\vec{q} - \vec{p}, \vec{p}) 
\hat{\varphi}_{A}(\vec{q} - \vec{p},\tau) \, \hat{\varphi}_{A}(\vec{p},\tau),
\nonumber\\
&& C^{(A)}_{m n}(\vec{a}, \vec{b}) = a_{m} b_{n} - \frac{\vec{a}\cdot\vec{b}}{3} \,\delta_{m n}, \qquad \vec{a} = \vec{q} - \vec{p}, \qquad \vec{b} = \vec{p}.
\label{phon6}
\end{eqnarray}
In practice $\hat{\pi}^{(A)}_{ij}(\vec{q}, \tau)$ is the Fourier transform of 
\begin{equation}
\pi_{i j}^{(A)}(\vec{x},\tau) = - \frac{1}{a^2} \biggl[\partial_{i} \hat{\varphi}_A\partial_{j} \hat{\varphi}_A - \frac{1}{3} (\partial_{k} \hat{\varphi}_{A} \partial^{k} \hat{\varphi}_{A}) \delta_{i j} \biggr],
\label{phon1}
\end{equation}
which is defined from the canonical energy-momentum tensor of the phonon field. In Eq. (\ref{phon6}) the field operators are defined as
\begin{equation}
\hat{\varphi}_{A}(\vec{k},\tau) = \hat{b}_{\vec{k}\, A} \, F_{k}(\tau) + \hat{b}_{-\vec{k}\, A}^{\dagger} F_{k,\,A}^{*}(\tau), \qquad
F_{k\,A}''+ 2 {\mathcal H} F_{k\,A}' + \biggl[ k^2 + 6 \alpha \frac{a''}{a} \biggr] F_{k\,A} =0.
\label{phon4}
\end{equation}
Eq. (\ref{phon4}) is derived 
by using a phonon action containing also the coupling to the Ricci scalar \cite{birrel}, i.e. 
\begin{equation}
S = \frac{1}{2}\int \sqrt{-g} \, d^{4} x\,\sum_{A = 1}^{{\mathcal N}}\, \biggl[ g^{\alpha\beta} \partial_{\alpha} \varphi_{A} \partial_{\beta} \varphi_{A} - \alpha\, R \, \varphi_{A}^2 \biggr],
\label{PHAC}
\end{equation}
where $R$ denotes the Ricci scalar.
If $\alpha =1/6$ the coupling is conformal; if $\alpha\neq 1/6$ the coupling is non-conformal. We shall assume, in agreement with Eq. (\ref{phon1}), 
  that the ${\mathcal N}$ scalars are minimally coupled (i.e. $\alpha =0$). Still it is useful to look at  the general form of Eq. (\ref{phon4}) whose general solution, in the 
case of a quasi-de Sitter (and spatially flat) background, is given in terms of Bessel functions. The 
dependence on a generic $\alpha$ can be easily included in the discussion and it amounts to a different Bessel index characterizing the evolution of the mode functions $F_{k, A}$. Along the same line, there
is  the possibility of including a mass term for the various phonons and this will result in a term $m_{A}^2 a^2$ inside the squared brackets of Eq. (\ref{phon4}). During a quasi-de Sitter stage of expansion 
the mass contribution introduces a further modification of the Bessel index since $m_{A}^2 a^2 \simeq (m_{A}^2/H^2) \tau^{-2}$. 
The fields $\varphi_{A}$ are fully inhomogeneous and do not have any background 
component. Their masses, if present, are assumed to be smaller than the inflationary curvature 
scale, as it will be discussed below. The set-up defined by Eq. (\ref{PHAC}) 
is not exhaustive and it is only aimed at illustrating, in a simplified situation, how 
secondary graviton spectra can be produced and enhanced by stimulated emission.

Using Eq. (\ref{phon6}) into Eq. (\ref{anis1b}) we can get a more explicit form of $\hat{\Pi}_{\oplus}(\vec{q},\tau)$ and 
$\hat{\Pi}_{\otimes}(\vec{q},\tau)$:
\begin{eqnarray}
&&\hat{\Pi}_{\oplus}(\vec{q},\tau) = \frac{1}{2 (2\pi)^{3/2}\, a^2(\tau)} \, \sum_{A = 1}^{{\mathcal N}}\, 
\int d^{3} p\, C_{\oplus}(\vec{q} - \vec{p}, \vec{p}) 
\hat{\varphi}_{A}(\vec{q} - \vec{p},\tau) \, \hat{\varphi}_{A}(\vec{p},\tau), 
\nonumber\\
&&\hat{\Pi}_{\otimes}(\vec{q},\tau) = \frac{1}{2 (2\pi)^{3/2}\, a^2(\tau)} \, \sum_{A = 1}^{{\mathcal N}}\, 
\int d^{3} p\, C_{\otimes}(\vec{q} - \vec{p}, \vec{p}) 
\hat{\varphi}_{A}(\vec{q} - \vec{p},\tau) \, \hat{\varphi}_{A}(\vec{p},\tau), 
\label{phon6a}
\end{eqnarray}
where, from Eq. (\ref{ST0}),
\begin{equation}
C_{\otimes}(\vec{a}, \vec{b}) = (\hat{m}\cdot\vec{a})(\hat{m}\cdot\vec{b}) - (\hat{n}\cdot\vec{a})(\hat{n}\cdot\vec{b}),\qquad C_{\oplus}(\vec{a}, \vec{b}) = (\hat{m}\cdot\vec{a})(\hat{n}\cdot\vec{b}) + (\hat{n}\cdot\vec{a})(\hat{m}\cdot\vec{b}).
\label{phon6b}
\end{equation}
Since, in Eq. (\ref{phon6a}), $\vec{a}= \vec{q} - \vec{p}$ and $\vec{b} = \vec{p}$ all the scalar products 
$\hat{m}\cdot\vec{q}$ and $\hat{n}\cdot\vec{q}$ vanish  given that $\hat{m}$, $\hat{n}$ and $\hat{q}$ are all orthogonal.
As a consequence the final expressions can be simplified even further:
\begin{equation}
C_{\otimes}(\vec{q}- \vec{p}, \vec{p}) = - e^{m n}_{\otimes}(\hat{q}) \, p_{m} \, p_{n}, \qquad
C_{\oplus}(\vec{q}- \vec{p}, \vec{p}) = - e^{m n}_{\oplus}(\hat{q}) \, p_{m} \, p_{n}.
\label{phon6c}
\end{equation}
Going now back to Eq. (\ref{phon7a}) the evolution equation for each polarization can then be written as
\begin{eqnarray}
&& \hat{h}_{\lambda}'' + 2 {\mathcal H} \hat{h}_{\lambda}' + q^2 \hat{h}_{\lambda} = - 2 a^2 \ell_{\mathrm{P}}^2 \hat{\Pi}_{\lambda},
\nonumber\\
&& \hat{\Pi}_{\lambda}(\vec{q},\tau) = - \frac{1}{2 a^2 (2\pi)^{3/2}} \int d^{3} p \, e^{m n}_{\lambda}(\hat{q}) \, p_{m} \, p_{n} 
\hat{\varphi}(\vec{q} - \vec{p}, \tau)\, \hat{\varphi}(\vec{p}, \tau). 
\label{phon6d}
\end{eqnarray}
We now have two complementary choices. Either we go on by representing the anisotropic 
stress on the basis of the tensor polarizations, or we go back to the tensor indices. In the second case, using Eq. 
(\ref{phon6d}) we obtain:
\begin{equation}
\hat{\Pi}_{ij}(\vec{q},\tau) = \sum_{\lambda= \otimes,\, \oplus} \, e^{(\lambda)}_{i j}(\hat{q}) \, \hat{\Pi}_{\lambda}(\vec{q},\tau) = 
- \frac{1}{2 a^2 (2\pi)^{3/2}} \sum_{A =1}^{{\mathcal N}}\int d^{3} p \, {\mathcal S}_{i j}^{(A)}(\hat{q}) \, p^{m} \, p^{n} \hat{\varphi}_{A}(\vec{q} - \vec{p},\tau) 
\hat{\varphi}_{A}(\vec{p}, \tau), 
\label{phon6e}
\end{equation}
where the sum over the polarizations has been performed with the help of Eq. (\ref{ST0B}), i.e. 
\begin{eqnarray}
{\mathcal S}_{ij}(\vec{q},\, \vec{p}) &=& p^{m} \, p^{n} \,\sum_{\lambda} e_{ij}^{(\lambda)}(\hat{q}) \, e_{m n}^{(\lambda)}(\hat{q}) = 4\, {\mathcal S}_{i j m n}(\hat{q}) \, p^{m} \, p^{n}
\nonumber\\
&=& p^2 \biggl\{ - \biggl[ 1 - \frac{(\vec{q}\cdot \vec{p})^2}{q^2 p^2}\biggr] \delta_{ij}  + \biggl[ 1 + \frac{(\vec{q}\cdot \vec{p})^2}{q^2 p^2}\biggr] \frac{q_{i} q_{j}}{q^2} + 2 \frac{p_{i} p_{j}}{p^2} 
\nonumber\\
&-& 2 (p_{i} q_{j} + q_{i} p_{j}) \frac{(\vec{q}\cdot \vec{p})}{q^2 p^2}
\biggr\}.
\label{phon6f}
\end{eqnarray}
It is simple to check that, indeed, Eq. (\ref{phon6e}) is traceless and divergence-less $\hat{\Pi}_{i}^{i} = q_{i} \hat{\Pi}_{j}^{i} =0$. To avoid confusions, note that 
${\mathcal S}_{ij}(\vec{q}, \vec{p}) = 4 p^{m}\, p^{n}\, {\mathcal S}_{i j m n}(\hat{q})$ 
in agreement with the definition given right after Eq. (\ref{ST4}). 
\subsection{Secondary Spectra}
Inserting Eq. (\ref{phon6e})  into Eq. (\ref{phon7a}), the resulting expression can be 
simplified by introducing an appropriately rescaled anisotropic stress
\begin{eqnarray}
&& \hat{h}_{ij}'' + 2 {\mathcal H} \hat{h}_{ij} + q^2 \hat{h}_{ij} =  \ell_{\mathrm{P}}^2  \hat{{\mathcal Q}}_{ij}(\vec{q},\tau), \qquad \hat{{\mathcal Q}}_{ij}(\vec{q},\tau) = \sum_{A=1}^{{\mathcal N}} \hat{{\mathcal Q}}^{(A)}_{ij},
\label{phon7}\\
&&\hat{{\mathcal Q}}^{(A)}_{ij}(\vec{q},\tau) = \frac{1}{ (2\pi)^{3/2}} \int d^{3} p \, {\mathcal S}^{(A)}_{ij}(\vec{q},\, \vec{p}) \hat{\varphi}_{A}(\vec{q} - \vec{p}, \tau)\, \hat{\varphi}_{A}(\vec{p}, \tau).
\label{phon7N}
\end{eqnarray}
By rescaling $\hat{h}_{ij}$, Eq. (\ref{phon7}) is further modified as:
\begin{equation}
\hat{\mu}_{ij}'' + \biggl[ q^2 - \frac{a''}{a} \biggr] \hat{\mu}_{ij} = \ell_{\mathrm{P}}^2 a(\tau) \hat{{\mathcal Q}}_{ij}(\vec{q},\tau), \qquad \hat{\mu}_{ij}(\vec{q},\tau) = a(\tau)\, \hat{h}_{ij}(\vec{q},\tau).
\label{phon8}
\end{equation}
The solution of Eq. (\ref{phon8}) is 
\begin{equation}
\hat{\mu}_{ij}(\vec{q},\tau) = \hat{\mu}_{ij}^{(1)}(\vec{q},\tau) +  \ell_{\mathrm{P}}^2 \int_{\tau_{\mathrm{i}}}^{\tau} 
a(\xi) \, G_{q}(\xi,\,\tau) \, \hat{{\mathcal Q}}_{ij}(\vec{q},\xi)\, d\xi,
\label{phon9}
\end{equation}
where
\begin{equation}
G_{q}(\xi, \tau) = \frac{f_{q}(\xi)\, f_{q}^{*}(\tau) - f_{q}^{*}(\xi)\, f_{q}(\tau)}{W(\xi)}, \qquad
W(\xi) = f_{q}(\xi)\, g_{q}^{*}(\xi) - f_{q}^{*}(\xi)\, g_{q}(\xi) =i.
\label{phon11}
\end{equation}
In terms of $\hat{h}_{ij}(\vec{q},\tau)$ the solution of Eq. (\ref{phon9}) becomes:
\begin{equation}
\hat{h}_{ij}(\vec{q},\tau) = \hat{h}^{(1)}_{ij}(\vec{q},\tau)  +\ell_{\mathrm{P}}^2 \int_{\tau_{\mathrm{i}}}^{\tau}
\overline{G}_{q}(\xi, \tau) \,  \hat{{\mathcal Q}}_{ij}(\vec{q},\xi)\, d\xi,
\qquad \overline{G}_{q}(\xi, \tau) = \frac{a(\xi)}{a(\tau)} \, G_{q}(\xi, \tau).
\label{phon12}
\end{equation}
Equation (\ref{phon12}) leads to the total spectrum of the gravitons in full analogy with the calculation of the primary contribution reported in Eq. (\ref{ST6A}), i.e.
\begin{equation}
\langle \Psi | \hat{h}_{ij}(\vec{q}, \tau) \, \hat{h}_{ij}^{\dagger}(\vec{p}, \tau) |  \Psi\rangle = 
\frac{2\pi^2}{q^3} P_{\mathrm{T}}(q,\tau)
\, \delta^{(3)}(\vec{q} - \vec{p}), 
\label{NC1}
\end{equation}
where now $P_{\mathrm{T}}(q, \tau) = P_{\mathrm{T}}^{(1)}(q,\tau) + P_{\mathrm{T}}^{(2)}(q,\tau)$ denotes the 
total power spectrum written as the sum of the primary contribution obtained in Eq. (\ref{ST6A}) and of the secondary contribution 
$P_{\mathrm{T}}^{(2)}(q, \tau) \, \delta^{(3)} (\vec{q} - \vec{p})$ which is explicitly given by the following expression:
\begin{equation}
\frac{\, q^3\,\ell_{\mathrm{P}}^4}{2 \pi^2} \, \sum_{A = 1}^{{\mathcal N}} \sum_{B = 1}^{{\mathcal N}} \, 
\int _{\tau_{i}}^{\tau} d\xi_{1}\, \int_{\tau_{i}}^{\tau} d\xi_{2} \, \overline{G}_{q}(\xi_{1},\, \tau) \, \overline{G}_{p}(\xi_{2}, \tau) \, 
\langle \Psi | \hat{{\mathcal Q}}^{(A)}_{ij}(\vec{q}, \xi_{1}) \, \hat{{\mathcal Q}}^{(B)}_{ij}(-\vec{p}, \xi_{2})| \Psi\rangle.
\label{NC2}
\end{equation}
Note that in Eq. (\ref{NC2}) we used that $ \hat{{\mathcal Q}}^{\dagger}_{ij}(\vec{p}, \xi) = \hat{{\mathcal Q}}_{ij}(- \vec{p}, \xi)$. It must be also appreciated that, as a consequence of the normalization of the Wronskian 
in Eq. (\ref{phon11}), we have $G_{q}(\xi,\tau) = G_{q}^{*}(\xi,\tau)$. 

To conclude this section it is appropriate to mention that the relation 
holding between the primary spectrum and the spectral energy density 
$\Omega_{\mathrm{GW}}(q,\tau)$ can be obtained also in the situation 
where the secondary spectrum is present. In particular it can be 
shown that 
\begin{equation}
\Omega_{\mathrm{GW}}(q,\tau) = \frac{q^2}{12 \, {\mathcal H}^2} f(q,q_{\mathrm{eq}}) \biggl[P_{\mathrm{T}}^{(1)}(q,\tau) + P_{\mathrm{T}}^{(2)}(q,\tau)\biggr],
\label{NC3}
\end{equation}
where the power spectra are evaluated for typical scales larger 
than the Hubble radius (i.e. $q \tau \ll 1$) prior to matter radiation 
equality. The function $f(q,q_{\mathrm{eq}})$ parametrizes the transfer function of the tensor spectra across matter-radiation equality \cite{mg1}
\begin{equation}
f(q,\,q_{\mathrm{eq}}) =1 + c_{1}\biggl(\frac{\nu_{\mathrm{eq}}}{\nu}\biggr) + b_{1}\biggl(\frac{\nu_{\mathrm{eq}}}{\nu}\biggr)^2,\qquad c_{1}= 0.5238,\qquad b_{1}=0.3537,
\label{NC4}
\end{equation}
where 
\begin{equation}
\nu_{\mathrm{eq}} = \frac{k_{\mathrm{eq}}}{2 \pi} = 1.288 \times 10^{-17} \biggl(\frac{h_{0}^2 \Omega_{\mathrm{M}0}}{0.1334}\biggr) \biggl(\frac{h_{0}^2 \Omega_{\mathrm{R}0}}{4.15 \times 10^{-5}}\biggr)^{-1/2}\,\, \mathrm{Hz};
\label{NC5}
\end{equation}
$\Omega_{\mathrm{M}0}$ and $\Omega_{\mathrm{R}0}$ denote, respectively, the present critical fractions of non-relativistic matter and radiation. 
\renewcommand{\theequation}{3.\arabic{equation}}
\setcounter{equation}{0}
\section{Quantum correlations}
\label{sec3}
The secondary power spectrum of Eq. (\ref{NC2}) is determined by the expectation value 
\begin{equation}
{\mathcal G}(\vec{q}_{1}, \vec{q}_{2};\xi_{1}, \xi_{2}) =\langle \Psi | \hat{{\mathcal Q}}^{(A)}_{ij}(\vec{q}_{1},\xi_{1})\, \hat{{\mathcal Q}}^{(B)}_{ij}(-\vec{q}_{2},\xi_{2})| \Psi \rangle,
\label{CORR1}
\end{equation}
whose explicit form is given by the following expression:
\begin{equation}
{\mathcal G}(\vec{q}_{1}, \vec{q}_{2};\xi_{1}, \xi_{2}) = \int \frac{d^{3} p_{1}}{(2\pi)^{3/2}} \int \frac{d^{3} p_{2}}{(2\pi)^{3/2}} \, {\mathcal S}_{(A)}^{ij}(\vec{q}_1; \vec{p}_{1}) \, {\mathcal S}_{(B)}^{ij}(\vec{q}_{2}; \vec{p}_{2}) {\mathcal M}_{A\, B}(\vec{q}_1,\, \vec{p}_{1};\, \vec{q}_{2},\, \vec{p}_{2}).
\label{CORR2}
\end{equation}
The explicit expression of ${\mathcal M}_{A\, B}(\vec{q}_1,\, \vec{p}_{1};\, \vec{q}_{2},\, \vec{p}_{2})$ appearing in Eq. (\ref{CORR2}) is quartic in the phonon fields:
\begin{equation}
{\mathcal M}_{A\, B}(\vec{q}_1,\, \vec{p}_{1};\, \vec{q}_{2},\, \vec{p}_{2}) =\langle \Psi | \hat{\varphi}_{A}(\vec{q}_{1}-\vec{p}_{1},\, \xi_{1}) \, 
\hat{\varphi}_{A}(\vec{p}_{1}, \xi_{1}) \, \hat{\varphi}_{B}^{\dagger}( \vec{q}_{2} -\vec{p}_{2}, \xi_{2}) \hat{\varphi}_{B}^{\dagger}(\vec{p}_{2}, \xi_{2}) | \Psi \rangle, 
\label{CORR3}
\end{equation}
and $\hat{\varphi}_{C}(\vec{k},\tau) = \hat{b}_{\vec{k}\, C} F_{k\, C}(\tau) + \hat{b}_{- \vec{k}\, C}^{\dagger} F_{k\, C}^{*}(\tau)$. In Eqs. (\ref{CORR1}) and (\ref{CORR3}) the average over the quantum state $|\Psi \rangle$ determines the properties of the final spectrum. 

The primary graviton spectrum $P_{\mathrm{T}}^{(1)}(q,\tau)$ 
depends on the averaged multiplicity of the initial state. The secondary graviton spectrum  $P_{\mathrm{T}}^{(2)}(q,\tau)$ depends on the expectation value of two anisotropic stresses i.e. four field operators. In terms of the 
statistical properties of the initial state this means 
that secondary graviton spectra are not only sensitive 
to the average multiplicity of the initial distribution but also to its variance.  There is here an analogy with quantum optics \cite{QO} where first-order correlation effects (i.e. Young interferometry) are unable to define completely the statistical properties of the light source. Second-order correlation effects are instead probed with Hanbury-Brown-Twiss interferometry where, instead of correlating 
the field amplitudes the correlation is between field {\em intensities} which are themselves quadratic in the field amplitudes. Similarly, primary and secondary graviton spectra depend, respectively, on the average of field amplitudes 
and field intensities.  In this connection, two issues need to be addressed: the problem of the ordering of the operators and the statistical properties of the initial quantum state. 

\subsection{Ordering of the operators}
Should we impose a specific ordering to the operators appearing in Eq. (\ref{CORR3})? 
If we do not impose any ordering,  the averages of the creation and annihilation operators appear in various combinations. Following the standard habit of quantum optics \cite{QO} and suppressing all the indices it is useful to consider the following normalized expectation values in a simple quantum mechanical framework where only one mode of the field is considered:
\begin{equation}
g^{(N)} = \frac{\langle \, \hat{b}^{\dagger}\, \hat{b}^{\dagger} \, \hat{b}\, \hat{b} \rangle}{\langle \, \hat{b}^{\dagger} \, \hat{b}\, \rangle^2}, \qquad 
g^{(A)} = \frac{\langle \, \hat{b}\, \hat{b}\, \hat{b}^{\dagger} \, \hat{b}^{\dagger}\, \rangle}{\langle \, \hat{b}^{\dagger} \, \hat{b}\, \rangle^2},
\qquad g^{(S)} = \frac{\langle \, \hat{b}\, \hat{b}^{\dagger} \, \hat{b}^{\dagger} \, \hat{b}\, \rangle}{\langle \, \hat{b}^{\dagger} \, \hat{b}\, \rangle^2},
\label{CORR4}
\end{equation}
where $[\hat{b},\, \hat{b}^{\dagger}] =1$. The superscripts of Eq. (\ref{CORR4}) denote, respectively, the normal, antinormal and symmetric ordering. The averages are  taken over a mixed state characterized by a density matrix
\begin{equation}
\hat{\rho} = \sum_{n= 0}^{\infty} {\mathcal P}_{n} \, |\, n\, \rangle \langle \, n\,|, \qquad \sum_{n= 0}^{\infty} {\mathcal P}_{n}  =1.
\label{CORR5}
\end{equation}
Using the commutation relations Eq. (\ref{CORR4}) can be written without mentioning any specific property of ${\mathcal P}_{n}$ as
\begin{equation}
g^{(A)} = g^{(N)} + \frac{3}{\langle \, \hat{n}\,  \rangle} + \frac{2}{\langle\, \hat{n}\,\rangle^2}, \qquad 
g^{(S)} = g^{(N)} + \frac{2}{\langle \, \hat{n}\,  \rangle},\qquad g^{(N)}= 1 + \frac{D^2 - \langle \, \hat{n}\,\rangle}{\langle \, \hat{n} \, \rangle^2}, 
\label{CORR6}
\end{equation}
where 
\begin{equation}
\langle\, \hat{n}\,\rangle = \mathrm{Tr}[\hat{\rho}\, \hat{b}^{\dagger}\, \hat{b}], \qquad 
\langle \, \hat{n}^2\, \rangle =  \mathrm{Tr}[\hat{\rho}\, \hat{b}^{\dagger}\, \hat{b} \, \hat{b}^{\dagger}\, \hat{b}],\qquad 
D^2 = \langle \, \hat{n}^2\, \rangle - \langle\, \hat{n}\,\rangle^2.
\label{CORR7}
\end{equation}
Consider now two examples associated with the density
matrix of Eq. (\ref{CORR5}): a Poisson mixture (denoted by ${\mathcal P}_{n}^{(P)}$) and a Bose-Einstein mixture (denoted by ${\mathcal P}_{n}^{(BE)}$); their explicit expressions are given, respectively, by:
\begin{equation}
{\mathcal P}_{n}^{(P)} = \frac{\overline{n}^{n} }{n \, !} e^{- \overline{n}}, \qquad {\mathcal P}_{n}^{(BE)} = \frac{1}{\overline{n} +1} \biggl( \frac{\overline{n}}{\overline{n} +1} \biggr)^{n}.
\label{CORR8}
\end{equation}
In the case of the Poisson mixture $g^{(N)} = 1$ since, in this case,  $\langle\, \hat{n}\,\rangle = \overline{n}$ and 
$D^2 = \overline{n}$ as it can be expected from the well known properties of the Poisson distribution. The 
antinormal and symmetric correlation functions $g^{(A)}$ and $g^{(S)}$ coincide, approximately, in the 
large $\overline{n}$ limit with the normal-ordered expression $g^{(N)}$.

In the case of the Bose-Einstein mixture $g^{(N)} = 2$ and, as in the Poisson case,  $g^{(A)} \simeq g^{(S)}$ for
$\overline{n} \gg 1$ since, from $P_{n}^{(BE)}$ of Eq. (\ref{CORR8}), $D^2 = \overline{n}^2 + \overline{n}$. Again the latter result is 
implied by known properties of the Bose-Einstein (or geometric) distribution. Let us finally consider the interesting example of the negative binomial mixture where 
the weight is given by:
\begin{equation}
{\mathcal P}_{n}^{(NB)} = \frac{\Gamma(\zeta + n)}{\Gamma(\zeta) \, \Gamma(n +1) } \, \biggl(\frac{\overline{n}}{\overline{n} + \zeta} \biggr)^{n} \, \biggl(\frac{\zeta}{\overline{n} + \zeta} \biggr)^{\zeta}.
\label{CORR9}
\end{equation}
Equation (\ref{CORR9}) corresponds to the distribution followed by $\zeta$ Bose-Einstein distributions 
all emitting at the same temperature. In the case of Eq. (\ref{CORR9})
\begin{equation}
g^{(N)} = 1 + \frac{1}{\zeta}, \qquad \frac{D^2}{\overline{n}^2} = 1 + \frac{1}{\zeta},
\label{CORR10}
\end{equation}
with $g^{(S)}$ and $g^{(A)}$ coinciding in the large $\overline{n}$ limit. The examples of Eqs. (\ref{CORR8})--(\ref{CORR10}) suggest two complementary conclusions.
The first conclusion is that since the normal-ordered expectation values dominate the normalized correlation functions, 
 it is useful to employ the normal ordering as long as the interesting physical limit is the one for $\overline{n} \gg 1$. The second conclusion is that the semiclassical approximation fails in the case of a Bose-Einstein initial state and, more 
generally, whenever $D^2 - \overline{n} \neq 0$. In quantum optics 
it is natural to impose the normal ordering in the higher-order correlators since the detection of light quanta (i.e. in the optical range of frequencies) occurs by measuring a photo-current, i.e. a current induced by the absorption of a photon \cite{QO}.
In the present case by normal-ordering the expectation values of two intensities (i.e. four field operators) we implicitly demand that  when the averaged multiplicity of the phonons goes to zero the secondary contribution to the graviton spectrum vanishes. This is consistent with the requirement that, ultimately, the interesting physical regime involves typical wavelengths much larger than the thermal wavelength where the averaged multiplicity 
of the (thermal) phonons is large.

\subsection{The density operator and the expectation values}
In the quantum mechanical examples discussed so far the dependence on the momenta has been neglected but their presence is essential. The expectation values are evaluated using the appropriate density operator. Recalling Eq. (\ref{eq2}) the density operator can be 
written as $\hat{\rho} = \hat{\rho}_{\mathrm{gravitons}} \otimes \hat{\rho}_{\mathrm{phonons}}$.
In the Fock basis the gravitons and the photons will then be 
characterized, in principle, by different statistical mixtures since the two density 
matrices span two different Hilbert spaces. The notations established 
in section \ref{sec2} (see, e.g.  Eqs. (\ref{ST2}) and (\ref{phon4}))  
imply that the gravitons are characterized by creation and annihilation 
operators conventionally denoted by 
$\hat{a}^{\dagger}_{\vec{k}\, \lambda}$ and $\hat{a}_{\vec{k}\, \lambda}$
while, for the phonons, the notation $\hat{b}^{\dagger}_{\vec{k}\, A}$ and $\hat{b}_{\vec{k}\, A}$ has been employed. The density matrix 
for the gravitons is therefore given by
\begin{equation}
\hat{\rho}_{\mathrm{gravitons}} = \sum_{\{m\}} \, {\mathcal P}^{(\mathrm{gravitons})}_{\{m\}} \, | \{m\} \rangle \langle \{m\}|,\qquad \sum_{\{m\}} \, {\mathcal P}^{(\mathrm{gravitons})}_{\{m\}} =1,
\label{MX1}
\end{equation}
where $ |\{m \}\rangle = |m_{\vec{k}_{1}\,\lambda} \rangle \, |  |m_{\vec{k}_{2}\,\lambda} \rangle \, |  |m_{\vec{k}_{3}\,\lambda} \rangle...$ and 
 the ellipses stand for all occupied modes of the field.  Similarly 
 the density matrix for the phonons can be written as: 
 \begin{equation}
\hat{\rho}_{\mathrm{phonons}} = \sum_{\{n\}} \, {\mathcal P}^{(\mathrm{phonons})}_{\{n\}} \, | \{n\} \rangle \langle \{n\}|,\qquad \sum_{\{n\}} \, {\mathcal P}^{(\mathrm{phonons})}_{\{n\}} =1.
\label{MX1a}
\end{equation}
The ${\mathcal P}_{\{n\}}$ (either for the phonons or for the gravitons) will be taken in the form\footnote{To avoid 
the pedantic repetition of the same formula with slightly different subscripts the indices denoting the polarizations of the gravitons (i.e. $\lambda$) and the different phonon fields (i.e. $A$) have been dropped.}: 
\begin{equation}
{\mathcal P}_{\{n\}} = \prod_{\vec{k}} \frac{\Gamma(\zeta_{\vec{k}} + n_{\vec{k}})}{\Gamma(\zeta_{\vec{k}}) \Gamma(n_{\vec{k}} +1)} 
\biggl(\frac{\overline{n}_{\vec{k}}}{\overline{n}_{\vec{k}} + \zeta_{\vec{k}}}\biggr)^{n_{\vec{k}}} \, \biggl(\frac{\zeta_{\vec{k}}}{\overline{n}_{\vec{k}} + \zeta_{\vec{k}}}\biggr)^{\zeta_{\vec{k}}},
\label{MX2}
\end{equation}
which is the field theoretical generalization of the quantum mechanical 
example of Eq. (\ref{CORR9}).
In the limit $\zeta_{\vec{k}} \to 1$ the distribution of Eq. (\ref{MX2}) 
becomes a Bose-Einstein distribution:
\begin{equation}
{\mathcal P}_{\{n\}} \to {\mathcal P}^{(BE)}_{\{n\}} = \prod_{\vec{k}} \frac{\overline{n}_{\vec{k}}^{n_{\vec{k}}}}{( 1 + \overline{n}_{\vec{k}})^{n_{\vec{k}} + 1 }}.
\label{MX3}
\end{equation}
Using Eqs. (\ref{MX1}) and (\ref{MX2}) the expectation values are computed with the help of the density operator:
\begin{equation}
\langle \hat{b}^{\dagger}_{\vec{q}} \, \hat{b}^{\dagger}_{\vec{p}}\, \hat{b}_{\vec{q}^{\,\prime}} \, \hat{b}_{\vec{p}^{\,\prime}} \rangle
= \mathrm{Tr} \biggl[ \hat{\rho}_{\mathrm{phonons}} \, \hat{b}^{\dagger}_{\vec{q}} \, \hat{b}^{\dagger}_{\vec{p}}\, \hat{b}_{\vec{q}^{\,\prime}} \, \hat{b}_{\vec{p}^{\,\prime}} \biggr].
\label{MX4}
\end{equation}
Using a stenographic notation\footnote{The operators $\hat{b}_{i}$ and $\hat{b}_{j}^{\dagger}$ denote the annihilation and creation operators related two generic momenta, 
i.e. for instance $\hat{b}_{\vec{q}}$ and $\hat{b}^{\dagger}_{\vec{p}}\,\,$; furthermore, following the same shorthand 
notation, $\delta_{i\, j}$ denotes the delta functions over the three-momenta (i.e. 
$ \delta_{\vec{q},\, \vec{p}}$).} where a each letter corresponds to a 
momentum the previous expectation value can be written as
\begin{eqnarray}
&& \langle \hat{b}_{i}^{\dagger} \,  \hat{b}_{j}^{\dagger} \, \hat{b}_{k}\,  \hat{b}_{\ell} \rangle = 
\langle \hat{b}_{i}^{\dagger} \,  \hat{b}_{i}^{\dagger} \, \hat{b}_{i}\,  \hat{b}_{i}\rangle \delta_{i\,j} \,
\delta_{j\,k}\, \delta_{\ell\,k} +
\nonumber\\
&& \langle \hat{b}_{i}^{\dagger} \,  \hat{b}_{j}^{\dagger} \, \hat{b}_{i}\,  \hat{b}_{j} \rangle  \, \delta_{i\, k} \,
 \delta_{j\, \ell} [ 1 - \delta_{ij}] +  \langle \hat{b}_{i}^{\dagger} \,  \hat{b}_{j}^{\dagger} \, \hat{b}_{j}\,  \hat{b}_{i} \rangle  \, \delta_{i\, \ell} \,
 \delta_{j\, k} [ 1 - \delta_{ij}].
\label{MX5}
\end{eqnarray}
In the term $\langle \hat{b}_{i}^{\dagger} \,  \hat{b}_{i}^{\dagger} \, \hat{b}_{i}\,  \hat{b}_{i}\rangle$ there are four different 
operators all acting on the {\em same} momentum; in the second class of terms (i.e. $ \langle \hat{b}_{i}^{\dagger} \,  \hat{b}_{j}^{\dagger} \, \hat{b}_{i}\,  \hat{b}_{j} \rangle$ and $\langle \hat{b}_{i}^{\dagger} \,  \hat{b}_{j}^{\dagger} \, \hat{b}_{j}\,  \hat{b}_{i} \rangle $) the momenta are paired two by two (as the corresponding deltas indicate). 
Using Eq. (\ref{MX2}) we shall have that 
\begin{eqnarray}
&& \langle \hat{b}_{i}^{\dagger} \,  \hat{b}_{i}^{\dagger} \, \hat{b}_{i}\,  \hat{b}_{i} \rangle = \sum_{m_{i}} m_{i} (m_{i} -1) P_{i}(m_{i}),\qquad 
\langle \hat{b}_{i}^{\dagger} \,  \hat{b}_{j}^{\dagger} \, \hat{b}_{i}\,  \hat{b}_{j} \rangle = \sum_{m_{i},\,\,m_{j}} m_{i}\,\, m_{j} \, P_{i\,j}(m_{i},m_{j}),
\nonumber\\
&& \langle \hat{b}_{i}^{\dagger} \,  \hat{b}_{j}^{\dagger} \, \hat{b}_{j}\,  \hat{b}_{i} \rangle = \sum_{m_{i},\,\,m_{j}} m_{i}\,\, m_{j} \, P_{i\,j}(m_{i},m_{j}),
\label{MX6}
\end{eqnarray}
where 
$P_{i}(m_{i}) = P_{k_{i}} (m_{k_{i}})$,  $P_{ij}(m_{i},m_{j})= P_{k_{i}} (m_{k_{i}})\,P_{k_{j}} (m_{k_{j}})$. The relations of Eq. (\ref{MX6}) follow
 from the definition of $P_{\{m\}}$, i.e. 
$P_{\{m\}} = P_{k_{1}}(m_{k_{1}})\,  P_{k_{2}}(m_{k_{2}})\, P_{k_{3}}(m_{k_{3}})\,....$ where the ellipses stand for the product over the various momenta.  Inserting Eq. (\ref{MX6}) into Eq. (\ref{MX5})  we get, after simple algebra, 
\begin{equation}
\langle \hat{b}_{i}^{\dagger} \,  \hat{b}_{j}^{\dagger} \, \hat{b}_{k}\,  \hat{b}_{\ell} \rangle = 
\overline{n}_{i}^2 \biggl(\frac{1}{\zeta_{i}}-1\biggr) \delta_{ij} \, \delta_{jk} \, \delta_{k\ell} + \overline{n}_{i} \, \overline{n}_{j} \biggl[ \delta_{ik} \delta_{j \ell} + 
\delta_{i \ell} \delta_{j k} \biggr].
\label{MX7}
\end{equation}
In the limit $\zeta_{i } \to 1$ for every mode of the field the contributions 
of the terms of the type $\langle \hat{b}_{i}^{\dagger} \,\hat{b}_{i}^{\dagger}\, \hat{b}_{i} \, \hat{b}_{i} \rangle $ exactly cancels; the standard rules of evaluating correlators in thermal field theory is quickly recovered \cite{kapusta}. If, however, $\zeta_{i} \neq 1$ the first contribution in Eq. (\ref{MX7}) does not vanish. 
The same procedure leading to Eq. (\ref{MX7}) can be used to compute, with the due differences, all the other expectation values arising when the ordering is either anti-normal or symmetric. Just to 
 give some examples consider, for instance,  the following results:
\begin{eqnarray}
\langle \hat{b}_{i}\,  \hat{b}_{j} \, \hat{b}_{k}^{\dagger} \,  \hat{b}_{\ell}^{\dagger} \rangle &=& \overline{n}_{i}^2 \biggl(\frac{1}{\zeta_{i}}-1\biggr) \delta_{ij} \, \delta_{jk} \, \delta_{k\,\ell} + (\overline{n}_{i}+1) \, (\overline{n}_{j} +1) \biggl[ \delta_{i\,k} \delta_{j \,\ell} + 
\delta_{i \,\ell} \delta_{j\, k} \biggr],
\label{MX8}\\
 \langle \hat{b}_{i}\,  \hat{b}_{j}^{\dagger} \, \hat{b}_{k}^{\dagger} \,  \hat{b}_{\ell}\rangle &=&  \overline{n}_{i}^2 \biggl(\frac{1}{\zeta_{i}}-1\biggr) \delta_{ij} \, \delta_{jk} \, \delta_{k\,\ell} + \overline{n}_{\ell} (\overline{n}_{i} + 1) \biggl[ \delta_{i\,j} \delta_{k\,\ell} + \delta_{i\, k}\delta_{\ell\, j}\biggr].
\label{MX9}
\end{eqnarray}
\subsection{Secondary spectrum}
Using the notations $\vec{k}_{1} = \vec{q}_{1} - \vec{p}_{1}$, and 
$\vec{k}_{2} = \vec{q}_{2} - \vec{p}_{2}$ the expectation value of Eq. (\ref{CORR3}) becomes
\begin{eqnarray}
&& {\mathcal M}_{A\, B}(\vec{q}_1,\, \vec{p}_{1};\, \vec{q}_{2},\, \vec{p}_{2})  =
 \langle \Psi |\hat{b}^{\dagger}_{-\vec{k}_{1}\, A} \, \hat{b}^{\dagger}_{-\vec{p}_{1}\, A}\, \hat{b}_{-\vec{k}_{2}\, B}\,  \hat{b}_{-\vec{p}_{2}\, B} | \Psi \rangle 
F_{k_{1}\,A}^{*}(\xi_{1}) F_{p_{1}\,A}^{*}(\xi_{1})F_{k_{2}\,B}(\xi_{2}) F_{p_{2}\,B}(\xi_{2}) 
\nonumber\\
&&+  \langle \Psi |\hat{b}^{\dagger}_{\vec{k}_{2}\,B} \, \hat{b}^{\dagger}_{\vec{p}_{2}\,B}\, \hat{b}_{\vec{k}_{1}\,A}\,  \hat{b}_{\vec{p}_{1}\, A} | \Psi \rangle 
F_{k_{2}\,B}^{*}(\xi_{2}) F_{p_{2}\,B}^{*}(\xi_{2})F_{k_{1}\,A}(\xi_{1}) F_{p_{1}\,A}(\xi_{1}) 
\nonumber\\
&&+  \langle \Psi |\hat{b}^{\dagger}_{-\vec{p}_{1}\,A} \, \hat{b}^{\dagger}_{\vec{k}_{2}\,B}\, \hat{b}_{\vec{k}_{1}\,A}\,  \hat{b}_{-\vec{p}_{2}\,B} | \Psi \rangle 
F_{k_{2}\,B}^{*}(\xi_{2}) F_{p_{2}\,B}(\xi_{2})F_{k_{1}\,A}(\xi_{1}) F^{*}_{p_{1}\,A}(\xi_{1}) 
\nonumber\\
&&+  \langle \Psi |\hat{b}^{\dagger}_{-\vec{p}_{1}\,A} \, \hat{b}^{\dagger}_{\vec{p}_{2}\,B}\, \hat{b}_{\vec{k}_{1}\,A}\,  \hat{b}_{-\vec{k}_{2}\,B} | \Psi \rangle 
F_{k_{2}\,B}(\xi_{2}) F^{*}_{p_{2}\,B}(\xi_{2})F_{k_{1}\,A}(\xi_{1}) F^{*}_{p_{1}\,A}(\xi_{1}) 
\nonumber\\
&&+  \langle \Psi |\hat{b}^{\dagger}_{-\vec{k}_{1}\,A} \, \hat{b}^{\dagger}_{\vec{k}_{2}\,B}\, \hat{b}_{\vec{p}_{1}\,A}\,  \hat{b}_{-\vec{p}_{2}\,B} | \Psi \rangle 
F_{k_{2}\,B}^{*}(\xi_{2}) F_{p_{2}\,B}(\xi_{2})F^{*}_{k_{1}\,A}(\xi_{1}) F_{p_{1}\,A}(\xi_{1})
\nonumber\\
&&+  \langle \Psi |\hat{b}^{\dagger}_{-\vec{k}_{1}\,A} \, \hat{b}^{\dagger}_{\vec{p}_{2}\,B}\, \hat{b}_{\vec{p}_{1}\,A}\,  \hat{b}_{-\vec{k}_{2}\,B} | \Psi \rangle 
F_{k_{2}\,B}(\xi_{2}) F^{*}_{p_{2}\,B}(\xi_{2})F^{*}_{k_{1}\,A}(\xi_{1}) F_{p_{1}\,A}(\xi_{1}).
\label{EX1}
\end{eqnarray}
Recalling that $[\hat{b}_{\vec{k}\, A}, \, \hat{b}^{\dagger}_{\vec{p}\, B}] = \delta^{(3)}(\vec{k} - \vec{p})\, \delta_{A\, B}$, the resulting expression for the secondary power spectrum becomes then:
\begin{eqnarray}
&& P_{\mathrm{T}}^{(2)}(q,\tau) = \frac{q^3 \,{\mathcal N}\,  \ell_{\mathrm{P}}^4}{4 \pi^5} \, \int d^{3} p \, 
\int_{\tau_{i}}^{\tau} d\xi_{1} \, \int_{\tau_{i}}^{\tau} d\xi_{2} 
\, \overline{G}_{q}(\xi_{1},\tau) \,   \overline{G}_{q}(\xi_{2},\tau)\, \overline{n}(|\vec{q} - \vec{p}|) \, \overline{n}(p) 
\nonumber\\
&& \times p^4\, \biggl[\frac{(\vec{q} \cdot\vec{p})^2}{q^2 p^2} - 1\biggr]^2 \, {\mathcal F}(|\vec{q} - \vec{p}|; \xi_{1}, \xi_{2}) \, {\mathcal F}(p; \xi_{1}, \xi_{2})
\nonumber\\
&&+  \frac{{\mathcal N}^2  \ell_{\mathrm{P}}^4}{64 \pi^5} \frac{q^7 \overline{n}^2(q)}{{\mathcal V}} \,\biggl(\frac{1}{\zeta_{q}} -1 \biggr)\int_{\tau_{i}}^{\tau} d\xi_{1} \,  \int_{\tau_{i}}^{\tau} d\xi_{2}\, \overline{G}_{q}(\xi_{1},\tau)\,  \, \overline{G}_{q}(\xi_{2},\tau) {\mathcal U}(q;\xi_{1}, \xi_{2}),
\label{EX2} 
\end{eqnarray}
where ${\mathcal V}$ denotes the fiducial normalization volume and the other functions are defined as
\begin{eqnarray}
&& {\mathcal F}(\vec{k};\, \xi_{1}, \xi_{2}) = F_{k}(\xi_{1}) \, F_{k}^{*}(\xi_{2}) 
+ F_{k}^{*}(\xi_{1}) \, F_{k}(\xi_{2}),
\label{EX3a}\\
&&{\mathcal U}(q;\xi_{1}, \xi_{2})= [F_{q}^2(\xi_{1})F_{q}^{*\,2}(\xi_{2}) +F_{q}^{*\,2}(\xi_{1})F_{q}^2(\xi_{2})].
\label{EX3}
\end{eqnarray}
The second term in Eq. (\ref{EX2}) vanishes in the Bose-Einstein limit (i.e. $\zeta_{q} \to 1$) but it vanishes also in the infinite volume limit.  This term is per se interesting 
but its analysis is not central to the present discussion. So, in what follows, we shall 
just assume that $\zeta_{q}=1$ which means that the phonons will be characterized 
by a single Bose-Einstein mixture. 
\renewcommand{\theequation}{4.\arabic{equation}}
\setcounter{equation}{0}
\section{Bose-Einstein enhancement}
\label{sec4}
To assess the Bose-Einstein enhancement of secondary graviton spectrum we must go back to 
Eqs. (\ref{phon11}) and (\ref{phon12}) and compute explicitly $G_{k}(\xi,\, \tau)$ and $\overline{G}_{k}(\xi,\,\tau)$.
For the estimates of the integrals appearing in the expressions of the secondary graviton spectra it will also 
be necessary to deduce more explicit forms of Eq. (\ref{EX3a}). Equations (\ref{ST1a}) and (\ref{phon4}) can be solved using standard techniques \cite{abr1,abr2,grad} in the case of quasi-de Sitter space-times.
\subsection{Explicit expression for the Green's functions}
From Eq. (\ref{phon11}) the explicit expression of $G_{q}(\xi,\tau)$ during a slow-roll phase is given by
\begin{equation}
G_{k}(\xi,\,\tau)= \frac{\pi}{2} \sqrt{\xi\tau} \biggl[ Y_{\nu}(- k \xi)\, J_{\nu}(- k \tau) - 
Y_{\nu}(- k \tau) \, J_{\nu}( - k \xi) \biggr], \qquad \nu = \frac{3 - \epsilon}{2 ( 1 - \epsilon)},
\label{BE1}
\end{equation}
where $\epsilon = - \dot{H}/H^2$ is the slow-roll parameter. Equation (\ref{BE1}) are derived by first solving Eq. (\ref{ST1a}) and the related equations for the tensor mode functions\footnote{The normalization $N$ appearing in Eqs. (\ref{BE2}) and (\ref{BE3}) should not be confused with 
${\mathcal N}$ counting the phonon fields.} 
\begin{eqnarray}
f_{\mathrm{i}}(k,\tau) &=& \frac{N}{\sqrt{2 k}} \sqrt{-k\tau} H^{(2)}_{\nu}(-\tau),\qquad N = \sqrt{\frac{\pi}{2}} e^{\frac{i}{2}(\nu + 1/2)\pi},
\label{BE2}\\
g_{\mathrm{i}}(k,\tau) &=& \partial_{\tau} f_{k}=  - N\sqrt{\frac{k}{2}} \sqrt{-k\tau}\biggl[ H^{(2)}_{\nu -1} (-x) +
\frac{(1 -2 \nu)}{2(-x)} H^{(2)}_{\nu} (-k\tau)\biggr] ,
\label{BE3}
\end{eqnarray}
where $H_{\nu}^{(1)}(z)$ is the Hankel functions of first kind \cite{abr1}. Notice that 
Eqs. (\ref{BE2}) and (\ref{BE3}) imply a constant Wronskian, i.e. $W(\xi) =i$. For power-law inflation we have $a(\tau) = (- \tau_{*}/\tau)^{-\beta}$ and $\nu = (\beta +1/2)$. In the latter case the rescaled Green functions are expressible as
\begin{equation}
\overline{G}_{k}(\xi,\tau) = - \frac{\pi}{2 k} (- k \xi) \biggl(\frac{\xi}{\tau}\biggr)^{- \nu} \biggl[ Y_{\nu}(- k \xi)\, J_{\nu}(- k \tau) - 
Y_{\nu}(- k \tau) \, J_{\nu}( - k \xi) \biggr].
\label{BE4}
\end{equation}
In the limit $\epsilon \ll 1$ we have that  $\overline{G}_{k}(\xi,\tau)$ is, approximately,
\begin{equation}
\overline{G}_{k}(\xi, \tau) = \frac{1}{k^3 \xi^2} \biggl\{ k (\xi - \tau) \cos{[k (\xi - \tau)]} - (k^2 \xi \tau + 1) \sin{[k (\xi - \tau)]}\biggr\}.
\label{BE5}
\end{equation}
It is also possible to deduce a more general expression for the Green's 
function valid across the transition between inflation and radiation. The expression for the mode function is 
\begin{eqnarray}
f_{k}(\tau) &=& \frac{1}{\sqrt{2 k}}\biggl\{\theta(-\tau - \tau_{*}) \biggl(1 - \frac{i}{k\tau}\biggr) e^{- i k\tau} 
\nonumber\\
&+& \theta(\tau+ \tau_{*}) \biggl[ \biggl(1 - \frac{i}{k\tau_{*}} - \frac{1}{2 k^2 \tau_{*}^2}\biggr) e^{ - i k\tau} + \frac{1}{2 k^2 \tau_{*}^2} e^{i k (\tau+ 2 \tau_{*})} \biggr] \biggr\},
\label{BE5a}
\end{eqnarray}
where $\tau_{*}$ denotes the transition time between the inflationary 
phase and the radiation-dominated phase and $\theta(z)$ denotes the Heaviside theta function with argument $z$. To derive Eq. (\ref{BE5a}) 
it is necessary to use an expression for the scale factor which is continuous 
across $-\tau_{*}$ such as 
\begin{equation}
a(\tau) = \biggl(\frac{-\tau_{*}}{\tau}\biggr) \theta(-\tau_{*} -\tau) + \biggl(\frac{\tau + 2 \tau_{*}}{\tau_{*}} \biggr) \theta(\tau+ \tau_{*}).
\label{BE5b}
\end{equation}
When the mode function is given by Eq. (\ref{BE5a}), the rescaled Green's function is
\begin{eqnarray}
\overline{G}_{q}(\xi,\,\tau) &=& \frac{1}{q} \biggl\{\theta(-\tau - \tau_{*}) 
\frac{q (\xi -\tau) \cos{[q (\xi - \tau)]} - (q^2 \xi \tau +1)\sin{[q (\xi - \tau)]}}{q^2 \xi^2} 
\nonumber\\
&-& \theta(\tau+ \tau_{*}) \biggl(\frac{\xi + 2 \tau_{*}}{\tau+ 2 \tau_{*}} \biggr) \, \sin{[q(\xi - \tau)]} \biggr\}.
\label{BE5c}
\end{eqnarray}
To estimate the secondary graviton spectrum  the functions defined in Eq. (\ref{EX3a}) must be computed; the function ${\mathcal F}(k;\, \xi_{1},\, \xi_{2})$ is given, in explicit terms, by
\begin{equation}
{\mathcal F}(k;\, \xi_{1},\, \xi_{2}) = \frac{\pi\, \sqrt{\xi_{1}\, \xi_{2}}}{2\, a(\xi_{1})\, a(\xi_{2})} \biggl[ J_{\mu}(- k \xi_{1}) \, J_{\mu}( - k \xi_{2}) + Y_{\mu}(- k \xi_{1}) Y_{\mu}(- k \xi_{2}) \biggr],
\label{BE8}
\end{equation}
where the Bessel index $\mu$ depends on the coupling of the phonons, i.e. 
$\mu \neq 1/2$ if the coupling is not conformal;  $\mu$ can also contain an explicit dependence on the mass of the phonon as already remarked in connection with Eq. (\ref{phon4}). For sake of simplicity  the mass of the various phonon species will be assumed to be always smaller than the de Sitter curvature scale. In the limit $k\xi_{1} \ll 1$ and $k\xi_{2} \ll 1$ the function ${\mathcal F}(k;\, \xi_{1},\, \xi_{2})$ can be expressed as:
\begin{equation}
{\mathcal F}(k;\, \xi_{1},\, \xi_{2}) \simeq \frac{\Gamma^2(\mu)}{k \pi a(\xi_{1}) a (\xi_{2})} \biggl( \frac{k^2 \xi_{1} \xi_{2}}{4} \biggr)^{1/2 - \mu} = \frac{4 \, \xi_{1}\, \xi_{2}}{a(\xi_{1})\, a(\xi_{2})} \, \frac{\Gamma^2(\mu)}{ k^3 \, \pi }  \biggl( \frac{k^2 \xi_{1} \xi_{2}}{4} \biggr)^{3/2 - \mu}.
\label{BE9}
\end{equation}
Instead of performing the integrations over $\xi_{1}$ and $\xi_{2}$ in Eq. (\ref{EX2}) it is useful to introduce the following set of rescaled variables
\begin{equation}
\alpha = \frac{p}{q},\qquad \beta = \frac{k}{q},\qquad 
x_{1} = q\,\xi_{1},\qquad x_{2} = q\,\xi_{2},\qquad y = q \,\tau.
\label{BE7}
\end{equation}
For instance, with the rescalings of Eq. (\ref{BE7}) the explicit expressions for the functions $\overline{G}_{q}(\xi,\tau)$ 
and ${\mathcal F}(p; \xi_{1}, \xi_{2})$ are, for $\tau < - \tau_{*}$:
\begin{eqnarray}
\overline{G}_{q}(\xi_{1},\tau) &=& \overline{G}_{q}(x_1, y) = 
\frac{(x_1 - y) \cos{(x_1 - y)} - (x_1 y +1) \sin{(x_1 - y)}}{q \,x_{1}^2},
\nonumber\\
\overline{G}_{q}(\xi_{2},\tau) &=& \overline{G}_{q}(x_2, y) = 
\frac{(x_2 - y) \cos{(x_2 - y)} - (x_2 y +1) \sin{(x_2 - y)}}{q \,x_{2}^2},
\nonumber\\
{\mathcal F}(p_{1};\, \xi_{1},\, \xi_{2}) &=& \frac{(\alpha^2 x_{1} x_{2} +1)\cos{[\alpha (x_{1} - x_{2})]} - \alpha (x_{2} - x_{1}) \sin{[\alpha (x_{1} - x_{2})]}}{p \,a(\xi_{1}) \,a(\xi_{2})\, \alpha^2 \,x_{1} \,x_{2}},
\nonumber\\
{\mathcal F}(k_{1};\, \xi_{1},\, \xi_{2}) &=& \frac{(\beta^2 x_{1} x_{2} +1)\cos{[\beta (x_{1} - x_{2})]} - \beta (x_{2} - x_{1}) \sin{[\beta (x_{1} - x_{2})]}}{k \, a(\xi_{1}) \,a(\xi_{2}) \,\beta^2 \,x_{1} \,x_{2}},
\label{BE6}
\end{eqnarray}
where we remind that, according to the notations of Eq. (\ref{EX1}), $k = |\vec{q}- \vec{p}|$.
\subsection{Explicit expressions for the secondary spectrum}
With the explicit formulas derived so far, the secondary spectrum can be 
expressed as:
\begin{equation}
P^{(2)}_{\mathrm{T}}(q,\tau) = \frac{{\mathcal N} \ell_{\mathrm{P}}^4}{4\pi^5} \int d^{3} p \, \int_{-\tau_{0}}^{\tau} \, d\xi_{1}\,  \int_{-\tau_{0}}^{\tau} \, d\xi_{2} \,\,{\mathcal L}(q,\,p;\, \xi_{1},\, \xi_{2},\, \tau) \,\,{\mathcal R}(q,\, p),
\label{BE10}
\end{equation}
where the functions ${\mathcal L}(q,\,p;\, \xi_{1},\, \xi_{2},\, \tau)$ and 
${\mathcal R}(q,\,p)$ are:
\begin{eqnarray}
{\mathcal L}(q,\,p;\, \xi_{1},\, \xi_{2},\, \tau) &=& \overline{G}_{q}(\xi_{1},\,\tau) \, \overline{G}_{q}(\xi_{2},\,\tau)\, {\mathcal F}(|\vec{q} - \vec{p}|;\, \xi_{1},\, \xi_{2}) \, {\mathcal F}(p;\, \xi_{1},\, \xi_{2}),
\label{BE11}\\
{\mathcal R}(q,p) &=& q^3\,p^4\,  \overline{n}(|\vec{q} - \vec{p}|) \, \overline{n}(p)\, \biggl[ \frac{(\vec{q} \cdot\vec{p})^2}{q^2 p^2} -1\biggr]^2.
\label{BE12}
\end{eqnarray}
Let us now consider, as an example, the case of minimal coupling 
of the phonons leading to $\mu \simeq 3/2$. In this case the integrals 
over $\xi_{1}$ and $\xi_{2}$ 
can be performed exactly by breaking the integration for $\tau< -\tau_{*}$ 
and for $\tau > - \tau_{*}$:
\begin{equation}
\int_{-\tau_{0}}^{\tau} \, \, d \xi_{1}\,\, \int_{-\tau_{0}}^{\tau} \, \, d \xi_{2}\,\,
{\mathcal L}(q,\,p;\, \xi_{1},\, \xi_{2},\, \tau) = \frac{\overline{{\mathcal A}}^2}{4} 
\frac{{\mathcal Z}^2(\tau,\, \tau_{0},\, \tau_{*})}{p^3\, q^4\, |\vec{q} - \vec{p}|^3},
\label{BE13}
\end{equation}
where the function ${\mathcal Z}(\tau,\, \tau_{0},\, \tau_{*})$ is
\begin{eqnarray}
&&{\mathcal Z}(\tau,\, \tau_{0},\, \tau_{*}) =\frac{q \tau_{0} \cos{[q(\tau + \tau_{0})]} - \sin{[q (\tau + \tau_{0})] }}{q \tau_{0}} + \frac{\sin{[q (\tau + \tau_{*})]} - q \tau \cos{[q (\tau + \tau_{*})]} }{q \tau_{*}}
\nonumber\\
&&+ \frac{ q \tau_{*} \cos{[ q (\tau + \tau_{*})]} + \sin{[q (\tau+\tau_{*})]}
- q (\tau+ 2 \tau_{*}) }{q (\tau+ 2 \tau_{*})}.
\label{BE14}
\end{eqnarray}
The amplitude of each intensity (quadratic in the field operators) has been denoted by $\overline{{\mathcal A}} = {\mathcal A}/\ell_{\mathrm{P}}^2$; from the latter 
relation it also follows that ${\mathcal A}= \ell_{\mathrm{P}}^2 \overline{{\mathcal A}}$ is the dimensionless amplitude given in Planck units. 
The correct physical limit where the function ${\mathcal Z}(\tau, \tau_{0}, \tau_{*})$ should be evaluated is the one where $q\tau \ll 1$, $q\tau_{*}\ll 1$ and $|\tau_{0}| \gg |\tau_{*}|$. Recall, indeed, that $\tau_{i} = - \tau_{0}$ corresponds to the initial integration time and $-\tau_{*}$ marks, 
in the sudden approximation, to the end of the quasi-de Sitter phase. 
After this limit the dependence on $\tau$ drops, to leading order.
Bearing then in mind the expression of ${\mathcal R}(q,p)$ of Eq. (\ref{BE12}) and putting 
all together we have that the expression of the secondary spectrum 
can be written as:
\begin{equation}
P_{\mathrm{T}}^{(2)}(q) = \frac{{\mathcal A}^2 {\mathcal N}}{16 \pi^5} 
\int \, p\,d^{3} p \, \frac{\overline{n}(|\vec{q} - \vec{p}|)\, \overline{n}(p)}{q \, |\vec{q} - \vec{p}|^3} \biggl[ \frac{(\vec{q} \cdot\vec{p})^2}{q^2 p^2} -1\biggr]^2.
\label{BE15}
\end{equation}
The dependence on the conformal time coordinate has been dropped since we are considering here the secondary contribution for the modes which are larger than the Hubble radius prior to matter-radiation equality. This is the quantity which must be directly compared with the standard form 
of the primary spectrum (see section \ref{sec5}).

In Eq. (\ref{BE15}) ${\mathcal A}$ denotes the common amplitude of the phonon fields; ${\mathcal A}$ depends on the Hubble rate in Planck units during the quasi-de Sitter stage of expansion and will be specified later 
for notational convenience. Without fine-tuning ${\mathcal A} \sim {\mathcal A}_{{\mathcal R}}$ where ${\mathcal A}_{{\mathcal R}}$ is the amplitude 
of curvature perturbations (see also section \ref{sec5}). The most interesting limit for the present analysis is the one where  $\overline{n}(p)\gg 1$ and $\overline{n}(|\vec{q} - \vec{p}|) \gg 1$. Thus $\overline{n}(p)$ and $\overline{n}(|\vec{q} - \vec{p}|)$ 
will be approximated as
\begin{equation}
\overline{n}(q) = \frac{1}{e^{q/q_{\mathrm{T}}} -1} \simeq \frac{q_{\mathrm{T}}}{q} , \qquad q < q_{\mathrm{T}},
\label{BE15a}
\end{equation}
where $q_{\mathrm{T}} = T$ is the (comoving) thermal wavenumber; we shall also assume that $\overline{n}(q)=0$ for $q>q_{\mathrm{T}}$. 
Using Eq. (\ref{BE15a}) inside Eq. (\ref{BE15}) the secondary spectrum can be written as
\begin{equation}
P_{\mathrm{T}}^{(2)}(q) = \frac{{\mathcal A}^2 {\mathcal N}}{16 \pi^5} \, \frac{q_{\mathrm{T}}^2}{q} \, {\mathcal I}(q,q_{\mathrm{T}}, k_{\mathrm{p}}), 
\label{BE15b}
\end{equation}
where $k_{\mathrm{p}}=0.002\,\mathrm{Mpc}^{-1}$ denotes the pivot scale already mentioned in the introduction and corresponding to a pivot frequency $\nu_{\mathrm{p}} = k_{\mathrm{p}}/(2\pi) =3.092 \times 10^{-18}\, \mathrm{Hz}$. The integral appearing in Eq. (\ref{BE15b}) is:
\begin{equation}
 {\mathcal I}(q,q_{\mathrm{T}}, k_{\mathrm{p}}) = \int  \frac{ d^{3} p}{ |\vec{q} - \vec{p}|^4}\biggl[ \frac{(\vec{q} \cdot\vec{p})^2}{q^2 p^2} -1\biggr]^2.
 \label{BE15c}
\end{equation}
By making explicit the angular and radial integrations and we shall have that 
\begin{equation}
{\mathcal I}(q,q_{\mathrm{T}}, k_{\mathrm{p}}) = \int_{0}^{2\pi} d\phi\, \int_{-1}^{1} d x\, \int_{k_{\mathrm{p}}}^{q_{\mathrm{T}}} 
p^2\,d p \, \,  \frac{(x^2 -1)^2}{(q^2 + p^2 - 2 q p x)^2},
\label{BE15d}
\end{equation}
where we used that $|\vec{q} - \vec{p}| = \sqrt{q^2 + p^2 - 2 q p x}$.
By explicitly performing the angular integrations in Eq. (\ref{BE15d}), the explicit expression 
of the secondary spectrum becomes
\begin{equation}
P_{\mathrm{T}}^{(2)}(q) = \frac{{\mathcal A}^2 {\mathcal N}}{192 \pi^4} 
\biggl(\frac{q}{q_{\mathrm{T}}}\biggr)^{-2}  \int_{k_{\mathrm{p}}}^{q_{\mathrm{T}}}
\frac{d p}{p} 
\biggl\{  \frac{4 \, [3 p^4 - 2 q^2 p^2 + 3 q^4]}{q^3\, p} 
+ 3 \frac{(p^2 - q^2)^2\,(p^2 + q^2)}{p^2 \, q^4} \ln{\biggl[\frac{|p - q|^2}{|p + q|^2}\biggr]}\biggr\}.
\label{BE17}
\end{equation}
To estimate the last angular integral appearing in Eq. (\ref{BE17}) we divide the integration interval 
in two regions, i.e. $k_{\mathrm{p}}\, \leq \, p < \, q$ and $q\, < p\, \leq q_{\mathrm{T}}$. In the first 
region we can expand the integrand in powers of $|q/p|<1$ and perform the integration over $p$. Similarly, in the 
second region the integrand can be expanded in powers of $|p/q| < 1$. The final result for the secondary spectrum for $k_{\mathrm{p}} \leq q < q_{\mathrm{T}}$ can then be expressed as:
\begin{equation}
P_{\mathrm{T}}^{(2)}(q) =\frac{52 \, {\mathcal N} \, {\mathcal A}^2}{315\, \pi^4} \biggl(\frac{q}{q_{\mathrm{T}}}\biggr)^{-2} \biggl[ 1 - \frac{7}{26} \biggl(\frac{k_{\mathrm{p}}}{q}\biggr)^3 - \frac{21}{26} \biggl(\frac{q}{q_{\mathrm{T}}}\biggr)+ \frac{1}{13} \biggl(\frac{q}{q_{\mathrm{T}}}\biggr)^3\biggr].
\label{BE18}
\end{equation}
Finally, as already remarked, for $q \gg q_{\mathrm{T}}$ the secondary spectrum 
is exponentially suppressed. 

\renewcommand{\theequation}{5.\arabic{equation}}
\setcounter{equation}{0}
\section{Concluding considerations}
\label{sec5}
Physical considerations suggest that the gravitons and the phonons have the same temperature. However different situations cannot be ruled out.
In what follows the primary and secondary contributions will be 
compared in different limits. By focussing on the primary contribution, let us insert Eq. (\ref{BE2}) and Eq. (\ref{BE15a}) into Eq. (\ref{ST6B}); in the limit $|q\,\tau|= |q/(a H)| \,\ll \, 1$ the result is:
\begin{equation}
P_{\mathrm{T}}^{(1)}(q) =\ell_{\mathrm{P}}^2 H^2 \frac{2^{2\nu}}{\pi^3} \Gamma^2(\nu) ( 1 - \epsilon)^{2\nu-1}  \biggl(\frac{q}{aH}\biggr)^{3 - 2 \nu} \coth{\biggl(\frac{q}{2 q_{\mathrm{T}}}\biggr)},
\label{TT1}
\end{equation}
which is nothing but the amplitude of the primary power spectrum
for typical length-scales larger than the Hubble radius prior to matter-radiation equality\footnote{ In Eq. (\ref{TT1}),  the term
$(1-\epsilon)^{2\nu -1}$ arises by eliminating $\tau$ is favor of $(aH)^{-1}$ since, within the slow-roll
approximation $a H = - 1/[ \tau ( 1 - \epsilon)]$.}. As already mentioned in Eq. (\ref{BE1}), $\nu= (3 - \epsilon)/[2 ( 1 - \epsilon)]$.
There are now different (but equivalent) ways of expressing the result of Eq. (\ref{TT1}).
Recalling that $\ell_{\mathrm{P}} = \overline{M}_{\mathrm{P}}^{-1}$ the spectrum 
(\ref{TT1}) at horizon crossing (i.e. $q \simeq H a$) becomes
\begin{equation}
P_{\mathrm{T}}^{(1)}(q)  = \frac{2}{3 \pi^2} \biggl(\frac{V}{\overline{M}_{\mathrm{P}}^4}\biggr)_{q\simeq a H}\, \coth{\biggl(\frac{q}{2 q_{\mathrm{T}}}\biggr)},
\label{TT2}
\end{equation}
where  the slow-roll relation $3 \overline{M}_{\mathrm{P}}^2 H^2 \simeq V$ has been used and $V$ denotes the inflaton potential. It is sometimes useful to express the normalization of Eq. (\ref{TT2}) in terms of $M_{\mathrm{P}} = \sqrt{8\pi } \,\,\overline{M}_{\mathrm{P}}$ with the result that the prefactor $[2/(3 \pi^2)] (V/\overline{M}_{\mathrm{P}}^4)$ appearing in Eq. (\ref{TT2}) is modified and becomes $(128/3)(V/M_{\mathrm{P}}^4)$.  From Eqs. (\ref{TT1}) and (\ref{TT3}), to first-order in the slow-roll approximation, the standard tensor spectral index $n_{\mathrm{T}}$ is
\begin{equation}
n_{\mathrm{T}} = 3 - 2 \nu \simeq - 2 \epsilon  + {\mathcal O}(\epsilon^2), \qquad r_{\mathrm{T}} = 16 \epsilon = - 8 n_{\mathrm{T}},
\label{TT3}
\end{equation}
where $r_{\mathrm{T}}$ is the ratio between the tensor and the scalar amplitude in the concordance model. The relation between the tensor amplitude 
and the inflaton potential  is not central to the present consideration so that the primary graviton spectrum can be usefully parametrized as
\begin{equation}
P^{(1)}_{\mathrm{T}}(q) = {\mathcal A}_{\mathrm{T}} \biggl(\frac{q}{k_{\mathrm{p}}}\biggr)^{n_{\mathrm{T}}} \,\coth{\biggl(\frac{q}{2 q_{\mathrm{T}}}\biggr)}, \qquad {\mathcal A}_{\mathrm{T}} = r_{\mathrm{T}} \, {\mathcal A}_{{\mathcal R}},
\label{TT4}
\end{equation}
where ${\mathcal A}_{\mathrm{T}} = r_{\mathrm{T}} {\mathcal A}_{{\mathcal R}}$ is the tensor amplitude expressed in terms of the amplitude 
of curvature perturbations ${\mathcal A}_{{\mathcal R}}$ at the pivot scale 
$k_{\mathrm{p}} =0.002\,\, \mathrm{Mpc}^{-1}$.  
The values of the cosmological parameters determined using the WMAP 7yr data alone \cite{wmap1,wmap2} in the light of the vanilla $\Lambda$CDM scenario are\footnote{Following the standard notations $\Omega_{\mathrm{c}}$, $\Omega_{\mathrm{b}}$ and $\Omega_{\mathrm{de}}$ are, respectively, the critical fractions 
of cold dark matter, baryons and dark energy;   
$\epsilon_{\mathrm{re}}$ denotes the optical depth at reionization and $n_{\mathrm{s}}$ is the spectral index 
of curvature perturbations; $h_{0}$ is the Hubble rate in units of $100\, \mathrm{km/[sec\, \times Mpc]}$.}: 
\begin{equation}
( \Omega_{\mathrm{b}}, \, \Omega_{\mathrm{c}}, \Omega_{\mathrm{de}},\, h_{0},\,n_{\mathrm{s}},\, \epsilon_{\mathrm{re}}) \equiv 
(0.0449,\, 0.222,\, 0.734,\,0.710,\, 0.963,\,0.088).
\label{Par1}
\end{equation}
If a tensor component is allowed in the analysis 
of the WMAP 7yr data alone the relevant cosmological parameters are determined to be \cite{wmap1,wmap2}
\begin{equation}
( \Omega_{\mathrm{b}}, \, \Omega_{\mathrm{c}}, \Omega_{\mathrm{de}},\, h_{0},\,n_{\mathrm{s}},\, \epsilon_{\mathrm{re}}) \equiv 
(0.0430,\, 0.200,\, 0.757,\,0.735,\, 0.982,\,0.091).
\label{Par2}
\end{equation}
In the case of Eq. (\ref{Par1}) the amplitude of the scalar modes is ${\mathcal A}_{{\mathcal R}} = 
(2.43 \pm 0.11) \times 10^{-9}$ while in the case of Eq. (\ref{Par2}) the corresponding values of ${\mathcal A}_{{\mathcal R}}$ and of $r_{\mathrm{T}}$ are given by 
\begin{equation}
{\mathcal A}_{{\mathcal R}} = (2.28 \pm 0.15)\times 10^{-9},\qquad r_{\mathrm{T}} < 0.36, 
\label{Par3}
\end{equation}
to $95$ \% confidence level. The qualitative features 
of the effects discussed here do not change if, for instance, one 
would endorse the parameters drawn from the minimal tensor extension 
of the $\Lambda$CDM paradigm and compared not to the WMAP 7yr 
data release but rather with the WMAP 3yr data release, implying, for instance, ${\mathcal A}_{{\mathcal R}} = 2.1^{+2.2}_{-2.3}\times 10^{-9}$, 
$n_{\mathrm{s}} =0.984$ and  $r_{\mathrm{T}} < 0.65$ (95 \% confidence level).

Let us now compare the primary and secondary contributions to the graviton spectrum in two different and 
extreme cases. In the first case the gravitons are not produced from stimulated emission but rather from spontaneous emission. This 
limit is recovered from Eq. (\ref{ST6B}) by setting $\overline{n}(q) \to 0$. Conversely 
the phonons are produced by stimulated emission. In the latter case the primary spectrum, unlike the secondary contribution, will not experience any Bose-Einstein 
enhancement. Consequently, the total power spectrum will be given by:
\begin{equation}
P_{\mathrm{T}}(q) \simeq {\mathcal A}_{\mathrm{T}} \biggl(\frac{q}{k_{\mathrm{p}}}\biggr)^{n_{\mathrm{T}}} + 
\frac{52 \, {\mathcal N} \, {\mathcal A}_{{\mathcal R}}^2}{315\, \pi^4} \biggl(\frac{q}{q_{\mathrm{T}}}\biggr)^{-2},
\label{Par4}
\end{equation}
where only the leading term of the secondary contribution has been kept (compare with Eq. (\ref{BE18})). In Eq. (\ref{Par4}) the amplitude 
of the phonon field has been estimated by setting ${\mathcal A} \simeq 
{\mathcal A}_{{\mathcal R}}$. This choice is fully justified since this is what happens in the case of a quasi-de Sitter phase.

To evaluate explicitly the figures in Eq. (\ref{Par4}) we still need an estimate 
of the ratio $q/q_{\mathrm{T}}$. Let us therefore proceed along this direction and remind the definition of $N_{\mathrm{max}}$, i.e. the maximal number of efolds which are today accessible by our observations:
\begin{equation}
e^{N_{\mathrm{max}}} = ( 2 \, \pi  \, \epsilon \, {\mathcal A}_{{\mathcal R}})^{1/4} \, ( \Omega_{\mathrm{R}0})^{1/4} \, \biggl(\frac{M_{\mathrm{P}}}{ H_{0} }\biggr)^{1/2} \biggl(\frac{H_{r}}{H}\biggr)^{\gamma -1/2},
\label{NN1}
\end{equation}
where $H_{0} = 100 \,h_{0} \,\mathrm{Mpc}^{-1}\, \mathrm{km}/\mathrm{sec}$ is the present value of the Hubble rate and $h_{0}^2 \Omega_{\mathrm{R}0} = 4.15 \times 10^{-5}$. In Eq. (\ref{NN1}) we included the possibility of a delayed reheating terminating at a putative scale $H_{r}$
possibly much smaller than the Hubble rate during inflation denoted by $H$; $\gamma$ controls the expansion rate during this intermediate phase.
Note that $H_{r}$ can be, at most, $10^{-44} M_{\mathrm{P}}$ 
corresponding to a reheating scale occurring just prior to the formation of the light nuclei. If $\gamma - 1/2 >0$ (as it happens if $\gamma = 2/3$ when the post-inflationary background is dominated by dust) $N_{\mathrm{max}}$ diminishes in comparison with the case 
when $H=H_{r}$.
Conversely if $\gamma - 1/2 <0$ (as it happens in $\gamma = 1/3$ 
when the post-inflationary background is dominated by stiff sources) $N_{\mathrm{max}}$ increases. If $H_{r} = H$ (or if $\gamma=1/2$) there is a sudden transition 
between the inflationary and the post-inflationary regimes and, in this case,
 we have approximately $N_{\mathrm{max}} \simeq 64 + 0.25 \ln{\epsilon}$. The value of $N_{\mathrm{max}}$ 
has then a large theoretical error, i.e., based on the previous considerations, $65\pm 15$. In practice $N_{\mathrm{max}}$ is determined by fitting 
the redshifted inflationary event horizon inside the present Hubble radius $H_{0}^{-1}$. If the total number of inflationary efolds $N_{\mathrm{t}}$ 
is larger than $N_{\mathrm{max}}$ (i.e. $N_{\mathrm{t}} > N_{\mathrm{max}}$), then the redshifted value of the inflationary event horizon 
is larger than the present value of the Hubble radius.
The value of the thermal wavenumber $q_{\mathrm{T}}$ is therefore 
given by
\begin{equation}
q_{\mathrm{T}} = H_{0} \, \biggl(\frac{45}{4 \pi^23 {\mathcal N}_{\mathrm{th}}}\biggr)^{1/4} \sqrt{\frac{M_{\mathrm{P}}}{H}} \, e^{ - (N_{\mathrm{t}} - N_{\mathrm{max}})},
\label{NN2}
\end{equation}
where ${\mathcal N}_{\mathrm{th}}$ denotes the total number of 
spin degrees of freedom in thermal equilibrium at the onset of inflation. 
From Eq. (\ref{NN2}) it is possible to argue for which 
values of the parameters $q < q_{\mathrm{T}}$ so that the possible 
enhancement of the power spectrum in Eq. (\ref{Par4}) is effectively realized. In this respect from Eq. (\ref{NN2}) it follows that
\begin{equation}
\frac{q}{q_{\mathrm{T}}} \simeq {\mathcal O}(10)\,\, e^{- ( N_{\mathrm{max}} - N_{\mathrm{t}})}\, \biggl(\frac{q}{h_{0}\mathrm{Mpc}^{-1}} \biggr)\, {\mathcal N}_{\mathrm{th}}^{1/4} \,\biggl(\frac{{\mathcal A}_{{\mathcal R}}}{2.20\times 10^{-9}}\biggr)^{1/4}
\biggl(\frac{\epsilon}{0.1}\biggr)^{1/4}.
\label{Par5}
\end{equation}
Choosing $N_{\mathrm{t}} \simeq N_{\mathrm{max}}$  and focusing the attention on the pivot scale (i.e. $q \simeq k_{\mathrm{p}}$ in Eq. (\ref{Par4})), the total 
spectrum of Eq. (\ref{Par4})  becomes:
\begin{equation}
P_{\mathrm{T}}(k_{\mathrm{p}}) \simeq {\mathcal A}_{\mathrm{T}} + 
\frac{52 \, {\mathcal N} \, {\mathcal A_{\mathcal R}}^2}{315\, \pi^4} \biggl(\frac{k_{\mathrm{p}}}{q_{\mathrm{T}}}\biggr)^{-2}.
\label{par7}
\end{equation}
Let us now suppose, in agreement with Eq. (\ref{Par5}), that $k_{\mathrm{p}}/q_{\mathrm{T}} \simeq {\mathcal O}(10^{-3})$ (at most) and ${\mathcal N} \simeq {\mathcal O}(10^{4})$ (at most). In the latter case the first term in Eq. (\ref{par7}) must be smaller than $10^{-10}$ (according to Eq. (\ref{Par3})) but the second term in Eq. (\ref{par7}) can be as large as ${\mathcal O}(10^{-13})$. 
Let us finally suppose that both the gravitons and the phonons are produced from stimulated emission. In this case Eq. (\ref{par7}) is modified as
  \begin{equation}
P_{\mathrm{T}}(q) \simeq {\mathcal A}_{\mathrm{T}} \biggl(\frac{q}{q_{\mathrm{T}}}\biggr)^{-1}\biggl[1 + 
\frac{52 \, {\mathcal N} \, {\mathcal A_{\mathcal R}}^2}{315\, \pi^4\, {\mathcal A}_{\mathrm{T}}} \biggl(\frac{q}{q_{\mathrm{T}}}\biggr)^{-1}\biggr],
\label{par8}
\end{equation}
The primary tensor spectrum is also enhanced so that the secondary contribution is always smaller except in the limit of a very large number of phonon fields such as ${\mathcal N} > {\mathcal O}(10^{4})$.  Note that Eq. (\ref{par8}) holds for $q < q_{\mathrm{T}}$; in the opposite  limit (i.e. 
$q > q_{\mathrm{T}}$) the secondary contribution vanishes while $P_{\mathrm{T}}(q) \simeq {\mathcal A}_{\mathrm{T}} (q/k_{\mathrm{p}})^{n_{\mathrm{T}}}$. If the total number of efolds is $ N_{\mathrm{t}} \sim N_{\mathrm{max}}$ it seems plausible that the primary and secondary contributions could be directly bounded by forthcoming satellite missions such as Planck explorer. The phenomenological signature will 
be a power spectrum with primary slope $(n_{\mathrm{T}} -1)$ and with secondary slope going as $-2$. 

The same hierarchies between the primary and secondary spectra are reflected, according to Eqs. (\ref{NC3})--(\ref{NC5}), in the spectral energy density $\Omega_{\mathrm{GW}}(q,\tau)$ whose explicit expression can be written, as a function of the frequency, as
\begin{eqnarray}
h_{0}^2 \Omega_{\mathrm{GW}}(\nu,\tau_{0}) &=& \Omega_{\mathrm{T}}\, f(\nu/\nu_{\mathrm{eq}}) \biggl[r_{\mathrm{T}} \, \biggl(\frac{\nu}{\nu_{\mathrm{p}}}\biggr)^{n_{\mathrm{T}}}\coth{\biggl(\frac{\nu}{2 \nu_{\mathrm{T}}}\biggr)} + \frac{52 \, {\mathcal N} \, {\mathcal A_{\mathcal R}}}{315\, \pi^4} \biggl(\frac{\nu}{\nu_{\mathrm{T}}}\biggr)^{-2}\biggr],
\label{WW1}\\
\Omega_{\mathrm{T}} &=& 3.94 \times 10^{-15} \biggl(\frac{h_{0}^2 \Omega_{\mathrm{R}0}}{4.15\times 10^{-5}}\biggr) \biggl(\frac{{\mathcal A}_{{\mathcal R}}}{2.28 \times 10^{-9}}\biggr).
\label{WW2}
\end{eqnarray}
For $\nu \geq \nu_{\mathrm{max}} \simeq \mathrm{GHz}$ $\Omega_{\mathrm{GW}}(\nu,\tau_{0})$ is exponentially suppressed\footnote{For the fiducial set 
of parameters adopted above we have that 
$\nu_{\mathrm{max}} = (\epsilon/0.1)^{1/4} [{\mathcal A}_{\mathcal R}/(2.28 \times 10^{-9})^{1/4}] [h_{0}^2 \Omega_{\mathrm{R}0}/(4.15 \times 10^{-5}]^{1/4}$ GHz. These frequencies will be however not touched by the present cosiderations.}
In spite of possible observational implications (which are notoriously difficult to assess in the case of the tensor modes of the geometry) the theoretical lesson we can draw from the present analysis is that the correlation properties of the initial quantum state can modify the tensor power spectrum at large-scales. This situation bears some analogy with the Glauber theory of optical coherence: first-order coherence of light is controlled by the analog of the primary 
graviton spectrum while second-order coherence is described by 
the analog of what has been called secondary power spectrum. 
Consequently, the secondary graviton spectra, in spite of their detectability prospects, 
reflect the statistical properties of  pre-inflationary mixed
states and of their second-order correlations.

\section*{Acknowledgements}
It is a pleasure to thank T. Basaglia and A. Gentil-Beccot of the 
CERN scientific information service for their kind assistance.


\newpage

\begin{thebibliography}{99}

\bibitem{wmap1} C.~L.~Bennett {\it et al.},  Astrophys.\ J.\ Suppl.\  {\bf 192}, 17 (2011);
N.~Jarosik {\it et al.},  Astrophys.\ J.\ Suppl.\  {\bf 192}, 14 (2011); 
 J.~L.~Weiland {\it et al.},  Astrophys.\ J.\ Suppl.\  {\bf 192}, 19 (2011).
 
 \bibitem{wmap2} D.~Larson {\it et al.}, Astrophys.\ J.\ Suppl.\  {\bf 192}, 16 (2011);
B.~Gold {\it et al.},  Astrophys.\ J.\ Suppl.\  {\bf 192}, 15 (2011);  
E.~Komatsu {\it et al.},   Astrophys.\ J.\ Suppl.\  {\bf 192}, 18 (2011).

\bibitem{ACBAR} C.~L.~Reichardt, P.~A.~R.~Ade, J.~J.~Bock {\it et al.}, Astrophys.\ J.\  {\bf 694}, 1200-1219 (2009).

\bibitem{QUAD} M.~Zemcov {\it et al.}  [QUaD collaboration], Astrophys.\ J.\  {\bf 710}, 1541 (2010); 
M.~L.~Brown {\it et al.}  [QUaD collaboration], Astrophys.\ J.\  {\bf 705}, 978 (2009).

\bibitem{LSS1} W.~J.~Percival, B.~A.~Reid, D.~J.~Eisenstein {\it et al.},  Mon.\ Not.\ Roy.\ Astron.\ Soc.\  {\bf 401}, 2148-2168 (2010).

\bibitem{LSS2} B.~A.~Reid, W.~J.~Percival, D.~J.~Eisenstein {\it et al.}, Mon.\ Not.\ Roy.\ Astron.\ Soc.\  {\bf 404}, 60-85 (2010).

\bibitem{SNN1} R.~Kessler, A.~Becker, D.~Cinabro {\it et al.},  Astrophys.\ J.\ Suppl.\  {\bf 185}, 32-84 (2009).

\bibitem{SNN2}  M.~Hicken, W.~M.~Wood-Vasey, S.~Blondin {\it et al.}, Astrophys.\ J.\  {\bf 700}, 1097-1140 (2009).

\bibitem{mg1} M.~Giovannini, Phys.\ Lett.\  {\bf B668}, 44 (2008); Class.\ Quant.\ Grav.\  {\bf 26}, 045004 (2009).

\bibitem{WB1} B.~Abbott {\it et al.}  [LIGO Collaboration],  Astrophys.\ J.\  {\bf 659}, 918 (2007); B.~Abbott {\it et al.}  [LIGO Collaboration], Phys.\ Rev.\  D {\bf 76}, 082003 (2007).
  
\bibitem{WB2}  B.~Abbott {\it et al.}  [ALLEGRO Collaboration and LIGO Scientific Collaboration],  Phys.\ Rev.\  D {\bf 76}, 022001 (2007).
 G.~Cella, C.~N.~Colacino, E.~Cuoco, A.~Di Virgilio, T.~Regimbau, E.~L.~Robinson and J.~T.~Whelan, Class.\ Quant.\ Grav.\  {\bf 24}, S639 (2007).  
  B.~P.~Abbott {\it et al.} [ LIGO Scientific and VIRGO Collaborations ], Nature {\bf 460}, 990 (2009).  
  
\bibitem{HF1} A.~Nishizawa, S.~Kawamura, T.~Akutsu {\it et al.},  Phys.\ Rev.\  {\bf D77}, 022002 (2008).

\bibitem{HF2} A. M. Cruise, Class. Quantum Grav. {\bf 17} , 2525 (2000); 
 A. M. Cruise and R. M. Ingley, Class. \ Quantum \ Grav. {\bf 23},  6185 (2006).

\bibitem{wein} S. Weinberg, {\it Cosmology}, (Oxford University Press, Oxford 2009).

\bibitem{one} P.~D.~B.~Collins and R.~F.~Langbein,ÊÊPhys.\ Rev.\ D {\bf 45}, 3429 (1992);  Phys.\ Rev.\ D {\bf 47}, 2302 (1993).

\bibitem{two} I.~Yu.~Sokolov, Class.\ Quant.\ Grav.\  {\bf 9}, L61 (1992).

\bibitem{twoa} M. Giovannini, Class. Quantum Grav. {\bf 29} 155003 (2012).

\bibitem{mg2}  M.~Gasperini, M.~Giovannini, and G.~Veneziano,  Phys.\ Rev.\  {\bf D48}, 439-443 (1993).

\bibitem{three} K.~Bhattacharya, S.~Mohanty and A.~Nautiyal,  Phys.\ Rev.\ Lett.\  {\bf 97}, 251301 (2006)

\bibitem{four} K.~Bhattacharya, S.~Mohanty and R.~Rangarajan, Phys.\ Rev.\ Lett.\  {\bf 96}, 121302 (2006).

\bibitem{five} W.~Zhao, D.~Baskaran and P.~Coles, Phys.\ Lett.\ B {\bf 680}, 411 (2009).

\bibitem{six}  M.~Giovannini,  Phys.\ Rev.\ D {\bf 83}, 023515 (2011).

\bibitem{seven} I.~Agullo and L.~Parker, ÊPhys.\ Rev.\ D {\bf 83}, 063526 (2011).
  
\bibitem{eight}   S.~Kundu, JCAP {\bf 1202}, 005 (2012).

\bibitem{birrel} N. D. Birrell and P. C. W. Davies, {\it Quantum fields in curved space}, 
(Cambridge University Press, Cambridge, UK, 1982).

\bibitem{ford} L. H. Ford and L. Parker, Phys. Rev. D {\bf 16},1601 (1977);   Phys. Rev. D {\bf 16}, 245  (1977).

\bibitem{landau1} L. D. Landau and E. M. Lifshitz, {\it The classical theory of Fields}
(Addison-Wesley and Pergamon Press, New York, 1971).

\bibitem{landau2} R. Isaacson, Phys. Rev. {\bf 166}, 1263 (1968); Phys. Rev. {\bf 166}, 1272 (1968).

\bibitem{landau3} L. R. Abramo, R. Brandenberger, and V. Mukahanov, 
Phys. Rev. D {\bf 56}, 3248 (1997); L. R. Abramo, Phys Rev. D {\bf 60}, 064004 (1999).

\bibitem{mg3} M.~Giovannini,  Phys.\ Rev.\ D {\bf 73}, 083505 (2006).

\bibitem{suzhang} D.~Su and Y.~Zhang,  Phys.\ Rev.\ D {\bf 85}, 104012 (2012).

\bibitem{QO} J.  Klauder and E. Sudarshan, {\it Fundamentals of quantum optics} (Benjamin, New York, 1968); 
 R. Loudon, {\it The quantum theory of light} (Clarendon Press, Oxford, 1983); 
  L. Mandel and E. Wolf, {\it Optical coherence and quantum optics}, (Cambridge University Press, Cambridge, 1995).

\bibitem{kapusta} J. Kapusta, {\it Finite-temperature field theory}, (Cambridge University Press, Cambridge 1989).

\bibitem{abr1}  M. Abramowitz and I. A. Stegun, {\it Handbook of Mathematical Functions} (Dover, New York, 1972).

\bibitem{abr2}  A. Erdelyi, W. Magnus, F. Obehettinger, and F. Tricomi,  {\it Higher Trascendental Functions} (Mc Graw-Hill, New York, 1953).  

\bibitem{grad} I. S. Gradshteyn and I. M. Ryzhik {\it Tables of Integrals, Series and Poducts}, (Academic Press, San Diego, 2000).


\end{thebibliography}
\end{document}